\documentclass[superscriptaddress,twocolumn,showpacs,preprintnumbers,amsmath,amssymb,aps]{revtex4}
\pdfoutput=1

\usepackage{amsmath,amssymb,amsthm,amsfonts}
\usepackage{MnSymbol} 
\usepackage{mathrsfs}
\usepackage{slashed}
\usepackage{graphicx}
\usepackage{subfigure}
\usepackage{color,rotating}
\usepackage{dsfont}
\usepackage{setspace}
\usepackage{verbatim}
\usepackage{hyperref}

\newcommand{\beq}{\begin{equation}}
\newcommand{\eeq}{\end{equation}}
\newcommand{\beqa}{\begin{eqnarray}}
\newcommand{\eeqa}{\end{eqnarray}}
\newcommand{\bfc}{\begin{figure}[t]\begin{center}}
\newcommand{\efc}{\end{center}\end{figure}}
\newcommand{\pd}{\partial}
\newcommand{\nn}{\nonumber}

\def\Fig#1{Fig.~\ref{#1}}
\def\fig#1{Fig.~\ref{#1}}
\def\Eq#1{Eq.~(\ref{#1})}
\def\eq#1{(\ref{#1})}
\def\sec#1{section~\ref{#1}}
\def\0#1#2{\frac{#1}{#2}}  

\graphicspath{{./figures/}}

\def\Z{\mathds{Z}}

\def\CC{{\mathcal C}}

\def\CO{{\mathcal O}}

\def\CV{{\mathcal V}}

\newcommand{\be}{\begin{eqnarray}}
\newcommand{\ee}{\end{eqnarray}}
\newcommand{\dm}{{\rm d}}

\begin{document}
\title{Functional renormalisation group in a finite volume}
\author{Leonard Fister} \affiliation{Institut de Physique Th\'eorique,
  CEA Saclay, F-91191 Gif-sur-Yvette, France.}

\author{Jan M. Pawlowski} \affiliation{Institut f\"ur Theoretische
  Physik, Universit\"at Heidelberg, Philosophenweg 16, 69120
  Heidelberg, Germany} \affiliation{ExtreMe Matter Institute EMMI, GSI
  Helmholtzzentrum f\"ur Schwerionenforschung mbH, Planckstr. 1,
  D-64291 Darmstadt, Germany}

\begin{abstract}
  We study a $\phi^4$-theory at finite temperature in a finite
  volume. Quantum, thermal and volume fluctuations are treated with
  the functional renormalisation group. Specifically, we focus on the
  interplay of temperature and length scales driving the system. We
  find that thermodynamical observables at finite volume such as the
  pressure approach the infinite volume limit similarly to that of the 
  vanishing temperature limit.

  We also advance the functional renormalisation group method at
  finite volume. In particular, we identify requirements for suitable
  regulators that admit the exponential thermal and finite volume
  decay properties.
\end{abstract}


\maketitle

\section{Introduction}
Finite volume/finite size effects play or may play an important
r$\hat{\rm o}$le in systems ranging from the physics of ultracold atom
clouds or optical lattices over condensed matter systems, multi-layer
systems to heavy-ion collisions where the interaction region is
relatively sharply bounded, though expanding. On the theoretical side,
lattice methods or other theoretical approaches with space-time coarse
graining are subject to finite volume effects due to finite number of
lattice or grid points.

In an Euclidean field theory finite volume effects are described
similarly to that of thermal fluctuations, hence being related to a
wealth of interesting phenomena and showing characteristic scaling
laws. In particular, systems with a second order phase transition lose
this property in a finite volume and show finite size
scaling. Prominent questions in this context concern the
approach to the infinite volume limit and the continuum limit in
lattice theories, the characteristics of the finite volume scaling
and the appropriate extraction of thermodynamical observables in a
given approach.

It is well-known from systems at finite temperature that thermal
effects both show a characteristic exponential decay with the masses
of the system at hand, as well as a characteristic polynomial decay
with momenta due to thermal contact terms. The latter contact terms
also play a pivotal r$\hat {\rm o}$le in thermodynamical
relations. Moreover, the strength of such contribution is only
computed with the correct exponential thermal decay. It is this
peculiar combination of exponential and polynomial suppression that
requires a quantitative control in non-perturbative approaches.

In the present work, we discuss finite temperature and finite volume
effects within the functional renormalisation group (FRG) approach,
see e.g.\
\cite{Braun:2005gy,Braun:2010vd,Braun:2011iz,Braun:2011uq,Tripolt:2013zfa}.
Here, we extend the discussions and approach set up in
Refs.~\cite{Fister:2011uw,Fister:2011um,Fister:Diss, Fister:2013bh,
  Fister:2013jrx} with a special emphasis on the interplay of the
scaling with temperature $T$, volume (box length $L$), and with the
infrared cutoff scale $k$ in the FRG approach.  This is worked out at
the example of a real $\phi^4$-theory. Only if all these scalings are
taken into account quantitatively one can expect to have access, in
particular, to thermodynamical observables and the equation of state.

A related functional approach is given by Dyson-Schwinger equations
(DSEs) at finite temperature and a finite volume, see
e.g.\ \cite{Fischer:2007pf,Luecker:2009bs,Bonnet:2011hh,Bonnet:2012az},
if the infrared cutoff scale $k$ is substituted with the ultraviolet
initial scale $\Lambda$, where the flow is initiated. The latter can be
identified with the ultraviolet momentum cutoff in the DSE framework.

The work is organised as follows. In Section~\ref{sec:FRG} we discuss
the FRG framework for $\phi^4$-theory,
including its properties and limits of the current simple
$\phi^4$-approximation. In Section~\ref{sec:finite} the formulation of
the FRG at finite temperature and volume is introduced in a
self-contained way, with a special emphasis on the fate of long range
correlations and the (non-)existence of condensates. In
Section~\ref{sec:thermodyn} we contrast thermodynamical quantities in
an infinite and finite volume, and their FRG evolution. Special
attention is paid to the relation of the chosen cutoff procedure
(regulator function) and the decay behaviour of thermal and finite
volume effects for large cutoff scales. In Section~\ref{sec:results}
we extend the decay analysis to the full dynamics. We also present
numerical results for the pressure, the dynamical mass and
$\phi^4$-coupling for finite volume and temperature adapted
regulators.

\section{$\phi^4$-theory and the Functional Renormalisation
  Group}\label{sec:FRG}

We study finite volume effects at the simple example of a
one-component scalar field theory. Its classical action is given by
\begin{align}%
\label{eq:phi4_action}
S[\phi] = \int \dm^4x \,\left[\012\left(\partial_\mu \phi(x)\right)^2+ V^{\rm
    cl}\left(\rho(x)\right)\right]\,, \quad  \rho
\equiv \012 \phi^2\,,
\end{align}
with the standard kinetic term and a potential
\begin{align}
\label{eq:pot}
V^{\rm cl}(\rho)= \0{\lambda}{2} (\rho-\kappa)^2
-\0{\lambda}{2} (\rho_0-\kappa)^2 \,, \quad
\rho_0=\max (0,\kappa)\,. 
\end{align}
The potential $V^{\rm cl}$ comprises the mass term and the
$\phi^4$-potential with coupling $\lambda$. The minimum of the
potential is at $\rho_0=\max(0\,,\,\kappa)$. The potential only
depends on $\phi^2$, thus the theory is $Z_2$-invariant under $\phi\to
-\phi$. The parameterisation of the potential in \eq{eq:pot} has the
normalisation $V^{\rm cl}(\rho_0)=0$ at the minimum $\rho_0$.  

\subsection{Functional renormalisation group}
The functional renormalisation group (FRG) is a functional continuum
approach that includes non-perturbative effects. In this Section we
highlight those aspects which are relevant for this work. 

The idea of the FRG is based on Wilson's idea of integrating out
fluctuations momentum-shell-wise. In practice, this is done by
starting at a microscopic scale $\Lambda$, where the theory is defined
via its classical action $S[\phi]$. 

In this work, we consider a field theory for a (one-component) scalar
$\phi$. Fluctuations from larger distance scales with respect to the
microscopic scale, i.e.\ energies lower than the renormalisation group
(RG) infrared scale $k$, are suppressed by adding a regulator term,
$R_k(p)$, which leaves the ultraviolet unchanged but serves as an
effective infrared mass $\sim k^2$. This allows to integrate
fluctuations which are of the order of this mass. By variation of the
scale from microscopic to macroscopic values fluctuations from all
scales can be taken into account successively. Within this procedure,
one smoothly interpolates between the classical action at large cutoff
scales $k$ and the full quantum effective action $\Gamma[\phi]$ at
vanishing cutoff scale, $k=0$. This is achieved in terms of an
integro-differential equation, the Wetterich equation,
\cite{Wetterich:1992yh},
\begin{align}
\label{eq:floweq}
\partial_t {\Gamma}_k[\phi] = \012 \int_q \, \01{\Gamma^{(2)}[\phi]+
  R_k}\partial_t {R}_k(q)\,, 
\end{align}
with the standard abbreviation $\int_q = \int \dm^4 q/(2\pi)^4 $, and
$t=\ln k/\Lambda$ with some reference scale $\Lambda$. The schematic
notation on the right hand side of \eq{eq:floweq} stands for
integration of the full, non-perturbative diagonal part of the
propagator $1/(\Gamma^{(2)}+R_k)(p,q)$ and the derivative of the
regulator $R_k(q)$ over the loop momentum $q$. The initial condition
for \eq{eq:floweq} is given by $\Gamma_{\Lambda}[\phi]=S[\phi]$. For
general $k$, the effective action $\Gamma_k[\phi]$ describes the full
quantum effective action for momentum scales larger than $k$, and
lacks the quantum fluctuations of momentum scale smaller than $k$. In
the limit $k\rightarrow 0$ it turns into the full quantum effective
action $\Gamma[\phi]$.
\bfc%
\includegraphics[width=\columnwidth]{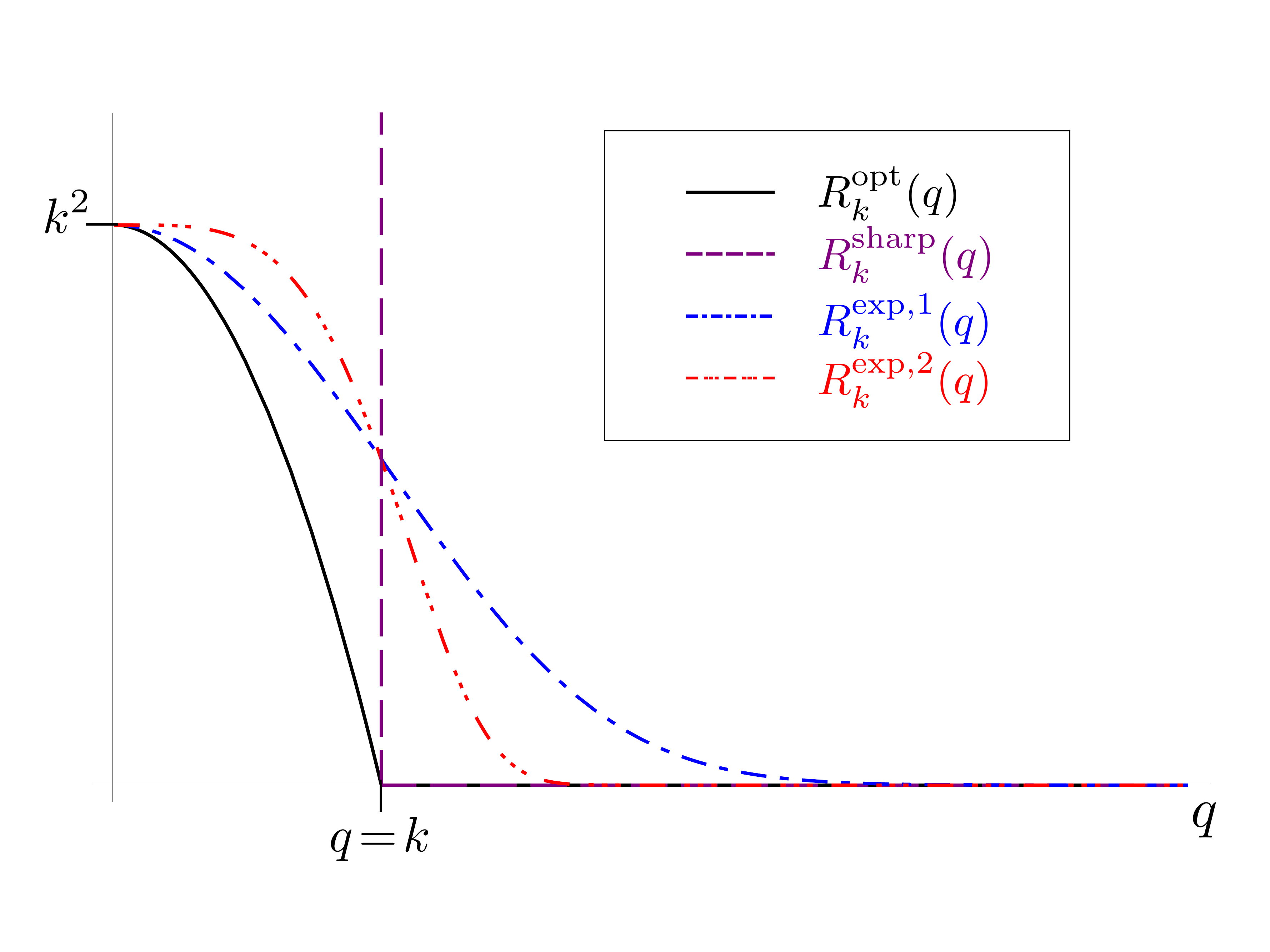}
\caption{Different regulator functions, see \eq{eq:Rexp}, \eq{eq:Rflat}
  and \eq{eq:Rsharp}.}
\label{fig:regulators}
\efc%

The regulator implements the shell-wise integration of fluctuations
described above, i.e.\ suppressing the infrared while leaving the
ultraviolet unmodified. Within these restrictions it can be chosen
freely. In the following, we study various common choices like the
exponential regulator,
\begin{align}%
R^{\rm exp,\,m}_{k}(q) = q^2 \,
\0{(q^2/k^2)^{m-1}}{e^{(q^2/k^2)^m}-1}\,,
\label{eq:Rexp}
\end{align}%
where the parameter $m$ controls the sharpness of suppression above
the scale $k$, the flat regulator \cite{Litim:2000ci},
\begin{align}%
R^{\text{\tiny{flat}}}_k(q) = \left(k^2-q^2\right) \Theta\left( k^2-q^2\right)\,,
\label{eq:Rflat}
\end{align}%
which is optimised within the lowest order of a derivative expansion,
see \cite{Litim:2000ci,Litim:2001up,Pawlowski:2005xe},
and the sharp regulator, \beq%
R^{\text{\tiny{sharp}}}_k(q) = k^2 \left(\0{1}{ \Theta\left(
      k^2-q^2\right)}-1\right)\,.
\label{eq:Rsharp}
\eeq%
The sharp regulator implements the standard ultraviolet momentum
regularisation within Dyson-Schwinger equations, as well as in
perturbation theory. There, loop momenta $q^2$ are cut at an
ultraviolet scale $\Lambda$ with $q^2\leq \Lambda^2$. Integrating the
flow with the sharp regulator from an ultraviolet initial scale
$k=\Lambda$ to $k=0$ precisely leads to such loop integrals: in this
case the flow gives the standard BPHZ-type renormalisation. Other
regulators give a generalised BPHZ-type renormalisation, see
\cite{Pawlowski:2005xe}, in particular the regulators \eq{eq:Rexp} do
not spoil the analytic properties of the integrands for finite
momenta. As we shall see, the non-analyticity of the flat regulator is
already causing large intricacies for the flow of thermodynamical
observables. These intricacies get enhanced by the stronger
non-analyticity of the sharp regulator.

The regulators \eq{eq:Rexp}, \eq{eq:Rflat}, \eq{eq:Rsharp} are shown in
\fig{fig:regulators}. Note that the regulator does not need to depend
on both, frequencies and spatial momenta. In fact, in the context of
thermal field theory it is convenient to regulate spatial momenta
only, $q^2\to \vec q^2$ in the regulators defined in
\eq{eq:Rexp}--\eq{eq:Rsharp}, because in this case Matsubara frequency
sums can often be performed analytically. We will elaborate on this in
\sec{sec:thermodyn}, where we focus also on the suitability of
regulators from additional constraints imposed by finite temperature
or volume.

\subsection{Local potential approximation and flow of the effective
  potential}

For large cutoff scales $k\to \Lambda$, the effective action
$\Gamma_k$ tends towards the classical action $S$ in
\eq{eq:phi4_action} with $\Lambda$-dependent parameters.  In
particular the effective potential $V_k(\rho)$ tends towards the
(bare) classical potential, $V_\Lambda=V_{\rm cl}$. Here, the
effective potential $V_k$ is defined as the effective action
$\Gamma_k$ evaluated for static (constant) fields $\rho_c$,
\begin{align}
\label{eq:Vk} V_k(\rho_c) \equiv \0{\Gamma_k [ \rho \! =\!
  \rho_c]}{\CV_{\text{\tiny{4d}} }}\,,\quad{\rm with}\quad
  \CV_{\text{\tiny{4d}}}=\prod_{i=\mu}^4\int_0^{L_\mu} d x_\mu\,.  
\end{align}
In \eq{eq:Vk} the trivial volume factor is removed. The effective
action also contains higher derivative terms which are dropped here in
the spirit of a low energy derivative expansion or local potential
approximation (LPA). In this approximation, the full effective action
reads
\begin{align}\label{eq:Gammak}
  \Gamma_k[\phi] = \int \dm^4x \left(\012\left( \partial_\mu
      \phi\right)^2+ V_k(\rho)\right)\,,
\end{align}
Within the LPA approximation \eq{eq:Gammak}, the $k$-dependent full
propagator is given by
\beq%
\label{eq:prop}
G_k(q,\rho) = \01{q^2 +m_k^2(\rho) + R_k(q)}\,,
\eeq
with the $k$- and field-dependent mass function  
\begin{align}\label{eq:mk}
\quad m^2_k(\rho) =
V_k'(\rho) + 2 \rho V_k''(\rho) \,. 
\end{align}
The primes indicate derivatives with respect to $\rho$. The effective
potential $V_k(\rho)$ can be expanded in powers of $(\rho-\rho_0)$
about the equation of motion (EoM) for static fields,
\begin{align}\label{eq:rho0} \left.\partial_\phi 
V\right|_{\phi=\sqrt{2\rho_0}}=0\,. 
\end{align} 
This leads to  
\begin{align}\nonumber 
  V_k(\rho) =&V_k(\rho_0) - \0{\lambda}{2} (\rho_0 -\kappa)^2+
  \0{\lambda}{2} (\rho -\kappa)^2  \\[2ex]
&+\sum_{n=3}^\infty
  \0{\lambda_n}{n!}(\rho-\rho_0)^n\,,
\label{eq:Vpol}\end{align}
with the relation between $\rho_0$ and $\kappa$ from \eq{eq:pot} and
$V_\Lambda(\rho_0)=0$. The first line comprises the normalisation of
the effective potential and the $k$-dependent counterpart of the
classical potential including fluctuation effects for $\kappa$ and
$\lambda$, while the second line comprises the fluctuation-induced
higher order scatterings.  In this expansion the $k$-dependent
analogues of the power-counting relevant classical parameters
$\lambda,\kappa$ are determined by
\begin{align}\label{eq:classpar} 
  \lambda_k(\rho_{0,k}-\kappa_k) = V'_k(\rho_{0,k})\,,\qquad
  \lambda_k=V''_k(\rho_{0,k}) \,,
\end{align}
where we made the $k$-dependence explicit. The mass function
$m_k^2(\rho)$, evaluated on the EoM, reads
\begin{align}\label{eq:exmass} 
m_k^2=m_k^2(\rho_{0,k})= \lambda_k(3 \rho_{0,k}-\kappa_k)\,, 
\end{align}
by using the definition of $\kappa$ in \eq{eq:pot}. Vanishing
$\kappa$ signals the phase transition. For large cutoff scales $k\to
\Lambda$, the effective potential tends towards the classical
potential $V_{\rm cl}$ given in \eq{eq:pot}. 

The flow equation for the effective potential is easily derived from
\eq{eq:floweq} and reads
\begin{align}\label{eq:dtV} 
\dot V_k(\rho)= \012 \int_q G(q,\rho) \dot R_k(q)\,, 
\end{align} 
with $G(q,\rho)$ defined in \eq{eq:prop}, and the dot indicates the
dimensionless derivative with respect to the RG scale, i.e.\ $\dot
V_k=\partial_t V_k$. The flow of the effective potential, \eq{eq:dtV}
depends on $V'$ and $V''$ via the mass function $m^2_k(\rho)$,
\eq{eq:mk} in the propagator. By taking the first and second
derivative of \eq{eq:dtV} with respect to $\rho$ in \eq{eq:phi4_action} we
find
\begin{align}\nonumber 
  \pd_\rho \dot{V}_k(\rho)
  &=   \012 \int_{q}  \0{\pd}{\pd \rho} G_k(q,\rho) \dot{R}_k(q)\,\nn\\[2ex]
  &= - \012 \left(m^2_k(\rho)\right)' \int_{q}\, G_k^2(q,\rho)
  \dot{R}_k(q)\,,
\label{eq:dtVp}\end{align}
and 
\begin{align}\nonumber 
\pd_\rho^2 \dot{V}_k(\rho) =&  \left[\left(m^2_k(\rho)\right)'\right]^2 \int_{q}
G_k^3(q,\rho) \pd_t R_k(q)\\[2ex]
& - \012 \left(m^2_k(\rho)\right)'' \int_{q}\, G_k^2(q,\rho) \dot{R}_k(q)\,.  
\label{eq:dtVpp}\end{align}
\Eq{eq:dtVp} and \eq{eq:dtVpp} depend on first and second
$\rho$-derivatives of the mass function, whose flows can be obtained
by further derivatives of the flow equation \eq{eq:dtVpp} with respect to
$\rho$.

\subsection{$\phi^4$-approximation}
The partial differential equation \eq{eq:dtV} can be solved within
various methods. Here, we resort to a Taylor expansion about the
minimum $\rho_0$. Then, evaluated at $\rho_0$,
\eq{eq:dtV}-\eq{eq:dtVpp} are the first three of an infinite hierarchy
of ordinary coupled differential equations for the couplings
$\kappa,\,\lambda,\, \lambda_{n\geq 3}$ defined in \eq{eq:Vpol}. At
the initial scale $\Lambda$ we have $\lambda_{n,\Lambda}=0$ for $n\geq
3$. Using the parameterisation \eq{eq:Vpol} we get for the left hand
side of the flow of the effective potential \eq{eq:dtV},
\begin{align}\nonumber 
  \partial_t V(\rho) = &\dot V(\rho_0) -\dot\rho_0
  \left[V'(\rho)-V'(\rho_0)\right]
  +\lambda (\dot \rho_0-\dot\kappa ) (\rho-\rho_0)\\[2ex]
\label{eq:flowlhs}
&+ \0{\dot \lambda}{2}(\rho-\kappa)^2-\0{\dot
  \lambda}{2}(\rho_0-\kappa)^2+\sum_{n\geq 3}
\0{\dot\lambda_{n}}{n!}(\rho-\rho_{0})^n\,,
\end{align}
leading to 
\begin{align}\label{eq:lhs} 
  \dot V'(\rho_0)= -\lambda \,\dot \kappa\,,\quad \dot V''(\rho_0)=
  \dot \lambda-\dot \rho_0\lambda_3\,.  
\end{align}
The $\lambda_3$ term in \eq{eq:lhs} signals the $k$-dependence of the
expansion point and feeds into the flow of $\lambda$. This exemplifies
the destabilising backreaction of the higher couplings on the flow of
the lower couplings in the presence of a flowing expansion point, and
can be avoided with a $k$-independent expansion point,
\cite{Pawlowski:2014zaa}. In the present work we drop the higher
couplings also for $k< \Lambda$, that is $\lambda_{n\geq 3}\equiv 0$
and approximate 
\begin{align}%
  V_k(\rho) \approx V_k^{(0)}(\rho_0) + \left(\rho-\rho_0\right)
  V_k'(\rho_0) + \012 \left(\rho-\rho_0\right)^2
  V_k''(\rho_0)\,, \end{align}%
with \eq{eq:classpar}. This is simply the first line in
\eq{eq:Vpol}. The rapid convergence of this expansion in $O(N)$-models
at infinite volume in general dimensions has been well-studied for
example within the computation of critical exponents, see e.g.\
\cite{Bohr:2000gp,Litim:2001dt,Litim:2002cf,Litim:2010tt}. This
approximation also works well for low-energy effective models for QCD
with mesonic degrees of freedom at vanishing density and finite
temperature \cite{Pawlowski:2014zaa}.  In turn, at finite density and
finite temperature higher powers in $\rho$, further couplings and
momentum dependences are required for quantitative statements, for
details see Ref.~\cite{Pawlowski:2014zaa}. This concerns in particular
the existence and location of a critical end point (CEP) in the phase
diagram. Then one either resorts to a $k$-independent expansion point
or relies on global techniques for solving partial differential
equations, see e.g.\ \cite{Braun:2010vd,Borchardt:2015rxa}.

Here, we are interested in structural results and we only keep the
relevant parameters $\kappa$ and $\lambda$ for the sake of
simplicity. Then the set of flow equations \eq{eq:dtV}-\eq{eq:dtVpp}
is closed. In this approximation, we have access to the flowing free
energy density $f_k$, the mass $m^2_k$, see \eq{eq:exmass}, and the
coupling $\lambda_k$ with
\begin{align}\label{eq:observablesf} 
  f_k\equiv V_k(\rho_0) = \0{\Gamma_k[\rho_0]}{\CV_{\text{\tiny{4d}}}}\,,
\end{align}
and 
\begin{align}\label{eq:observables} 
\ m_k^2=V'_k(\rho_0)+2 \rho_0
  V_k''(\rho_0)\,, \quad \ \lambda_k \equiv V_k''(\rho_0)\,.
\end{align}
The two parameters in \eq{eq:observables} can be determined from the 
coupled set of flow equations for $\partial_\rho\dot V_k (\rho_0)$ and
$\partial_\rho^2 \dot V_k(\rho_0)$. The flow of the free energy density
$f_k$ in \eq{eq:observablesf} only depends on $\kappa_k$ and
$\lambda_k$, and can be integrated separately with the solution for
$\kappa_k$ and $\lambda_k$.

In the current $\phi^4$-approximation the mass function and its
derivatives are given by
\begin{align}\label{eq:massphi4} 
  m_k^2(\rho) = \lambda_k(3 \rho-\kappa_k)\,,\quad {m^2_k}'=3
  \lambda_k\,, \quad {m^2_k}''\equiv 0\,.
\end{align} 
In summary, this leads to the closed system of flow equations for
$f_k, \kappa_k$ and $\lambda_k$, with 
\begin{align}\nonumber 
  \dot f_k &= \012 \int_q \dot{R}_k(q)\,G_k(q)\,,\\[2ex]\nonumber 
\dot  \kappa_k &= \032 
\int_q\dot{R}_k(q)\, G^2_k(q)\,,\\[2ex]
\label{eq:LPAobserve}
\dot{\lambda}_k &= 9\lambda_k^2 \int_q \dot{R}_k(q)\,G^3_k(q)\,.
\end{align}%
Note that all flows are positive and therefore leading to decreasing
functions $f_k,\kappa_k$ and $\lambda_k$. In summary this results in
an increasing mass. This reflects the property that bosonic flows are
symmetry-restoring.  Note also that the reduced system of equations
for the mass and the coupling is already closed. The flow of $f_k$
only depends on $m_k=\lambda_k |\kappa_k|$.  The above flows
integrations over the four-momentum $q$: for finite extent in either
temporal or spatial direction the integration (partly) turns into
summation over discrete modes. We elaborate on these modifications in
\sec{sec:finite}.

\section{Finite temperature \& volume}\label{sec:finite}

At finite temperature and/or in a finite volume, some of the
space-time directions only have finite extent,
\begin{align}\label{eq:finiteex}
x_0\in [0,1/T]\,,\quad {\rm  and/or}\quad 
x_i\in[0\,,\,L_i]\,. 
\end{align}
Note that in the current Euclidean formulation there is no technical
difference between the spatial and temporal directions.  One might as
well interpret a finite extent in the temporal direction as a theory
at vanishing temperature and one compact spatial direction with
$L_0=1/T$.  A finite extent in a given direction with periodic
boundary conditions (for bosons, $\phi(x+L_\mu)=\phi(x)$)) only allows
for plane waves that are periodic under shifts $x_\mu \to x_\mu +
L_\mu$, that are $\exp (i 2 \pi n x_\mu/L_\mu)$ with $n\in \Z$. This
entails that frequencies and momenta are discrete,
\begin{align}\label{eq:mat}
  p_\mu = 2 \pi n\0{1}{L_\mu}\,,\quad {\rm with}\quad n\in \Z\quad
  {\rm and}\quad T=\0{1}{L_0}\,.
\end{align}
In the case of finite temperature these are the Matsubara
frequencies. Accordingly, spatial momentum integrations turn into sums
over spatial Matsubara modes.  The compactification of a coordinate
$x$ related to the momentum $p$ to the interval $x\in [0\,,\, L]$ with
periodic boundary conditions leads to
\begin{align}%
  \01{2\pi}\int_{-\infty}^{\infty}\dm q\, I(q^2)
  \stackrel{L<\infty}\rightarrow \01L \sum_{n=-\infty}^{\infty}
  \,I\left(\left(\0{2\pi n}{L}\right)^2\right)\,.
\end{align}%
This is trivially extended to higher dimensions.

In the case of finite temperature the temporal direction $x_0$ is
compactified to the range $x_0\in [0,\, 1/T]$ with periodic boundary
conditions for bosons. For fermions one has anti-periodic boundary
conditions. For infinite spatial extent, momentum integrations have
the form
\begin{align}
  \int_{q}\,I(q^2) \ \ \stackrel{T>0}{\longrightarrow}\ \
  T\!\!\!\sum_{n_0=-\infty}^{\infty} \int_{\vec q} I\left((2\pi T
    n_0)^2 +\vec q^2\right)\,,
\label{eq:finiteTint}\end{align}
with the spatial momentum integration $\int_{\vec q}= \int \dm^3 q/(2
\pi)^3$, and a sum over the Matsubara modes $2\pi T n_0$.

In the case of finite temperature and finite spatial extent $L$ we are
left with sums over discrete modes in both temporal and spatial
directions, respectively, viz.
\begin{align}
& \int_{q}I(q^2) \xrightarrow[L<\infty]{T>0}
\nn\\[2ex]
& \phantom{aaa} \frac{T}{L^3} \sum_{\substack{n_\mu=-\infty \\
    \mu=0,1,2,3}}^{\infty} I\left(4\pi^2\left[T^2 n_0^2
    +\0{n_1^2+n_2^2+n_3^2}{L^2}\right]\right)\,,
\label{eq:loopfinV} \end{align}%
with $n_1$, $n_2$ and $n_3$ label the discrete spatial modes in the
cubic volume of edge length $L$.

\subsection{Range of finite temperature \& volume effects} \label{sec:range}

Finite extent, \eq{eq:finiteex}, seemingly induces a dimensional
reduction of the theory in the presence of finite temperature and/or
finite volume: for large temperature/small volume only the zero mode of
the sum contributes, leaving only the integration/sum over the
other momentum directions. This dimensional reduction is well-known at
finite temperature. There it is commonly formulated that a quantum
field theory in $3+1$ (or other dimensions) is dimensional reduced to
its $3$-dimensional counterpart for sufficiently large
temperature. However, strictly speaking this only holds for
momentum scales $p/T\ll 1$. There, however, it is valid for all
temperatures. In turn, for $p/T\gg 1$ the dimensional reduction does
not take place. This is reflected in the fact that the
renormalisation can be chosen to be temperature-independent.

The co-existence and interaction of these two momentum regimes for all
temperatures and/or finite lengths has the direct consequence that we
see both, an exponential suppression of thermal and/or finite volume effects
with the physical mass scales of the theory as well as the standard
polynomial suppression
of large momentum modes already present in perturbation theory: \\[-2ex]

\noindent {\it Exponential decay:} In finite 
temperature field theory thermal correlations decay exponentially
with the mass gap $m_{\rm gap}$ of the theory, which can be formalised
as
\begin{align}\label{eq:Tdecay} 
  \lim_{ \0{m_{\rm gap}}{T}\to \infty} \left|\0{\CO_n^T -
      \CO_n^{T=0}}{\CO_n^{T=0} }\right|\propto A\left(\0{m_{\rm
        gap}}{T}\right)\, \exp\left( -\0{m_{\rm gap} }{T}\right)\,,
\end{align}
for correlation functions  
\begin{align}
  \CO_{n}^T= \langle \phi(x_1)\cdots \phi(x_n)\rangle_{T}\,, 
\end{align}
in the absence of further scales. If the latter ones are present they lead to further
dependencies on their dimensionless ratios in $A$. The function $A(y)$
is a rational function of the argument $y=m_{\rm gap}/T$, and the exponential
suppression with $\exp(-m_{\rm gap}/T)$ can be readily computed from
one loop thermal perturbation theory. Evidently, similar expressions
hold for the finite volume correlations, with the identification $T\to
1/L_i$ in \eq{eq:Tdecay}. For potentially vanishing
correlation functions the denominator of \eq{eq:Tdecay} should be
substituted with $\left|\CO_n^T\right| + \left|
    \CO_n^{T=0}\right|$ for the sake of
definiteness. For the correlation functions studied in the present
work this is not necessary. \\[-2ex]

\noindent {\it Polynomial decay:} Note that \eq{eq:Tdecay} does {\it
  not} imply an exponential suppression of thermal corrections in
momentum space with momenta proportional to $\exp (-{\rm const.}\,
p/T)$. In momentum space, correlation functions show a subleading
momentum behaviour with temperature-dependent coefficients that is
only suppressed by powers of $p^2$. For the sake of simplicity, we
consider the symmetric point $p_i^2=p^2$ for all $i=1,...,n$ and
obtain
\begin{align}\label{eq:Tpoldecay} 
  \lim_{\0{p}{T}\to \infty}\left|\0{\CO_{n}^T - \CO_{n}^{T=0}}{\CO_{n}^{T=0}}\right|
  \propto B\left(\0{p}{T}\right)\to 0 \,, 
\end{align}
with a rational function $B(p/T)$ which decays polynomially for
$p/T\to\infty$. \\[-2ex]

For the two-point correlation functions, the propagator $G(p)$, this
is known as the Tan-contact term in e.g.\ the context of ultracold
atoms \cite{Tan:2008-1,Tan:2008-2,Tan:2008-3}. Within the FRG these
terms have been discussed in
\cite{Fister:2011uw,Fister:Diss,Boettcher:2012dh} for Yang-Mills
theory and ultracold atoms, respectively.  In
\cite{Fister:2011uw,Fister:2011um,Fister:Diss} a detailed analysis of the
diagrammatic origin of this behaviour was provided. This also entails
that the coefficients of this decay come from thermal fluctuations and
their values hinge crucially on the correct implementation of the
exponential decay in \eq{eq:Tpoldecay}. Hence, even though these terms
have a polynomial decay, the correct computation of the prefactor
$C_T$ relies on the exponential suppression \eq{eq:Tdecay} with the
physical mass scales of the theory. For the propagator we have
\begin{align}\label{eq:TpoldecayG} 
  \lim_{\0{|\vec p|}{T}\to 0}\left|\0{G^{T}(0,\vec p )- G^{T=0}(0,\vec
      p)}{G^{T=0}(0,\vec p)}\right| \propto C_T\left(\0{T}{|\vec
      p|}\right)^4 \,,
\end{align}
with a temperature-dependent prefactor $C_T$ (in the presence of other
scales), the Tan-contact. Importantly, the Tan-contact is related to
e.g.\ the equation of state as well as the density. Hence, it is
expected that only an approach that is able to deal with both the
exponential suppression and the polynomial terms enables us to compute
these quantities.

\subsection{Flow equation at finite temperature \&
  volume} \label{sec:flowTV}

The flow equation for the effective potential takes the form
\begin{align}\label{eq:flowTL}
\dot V_k(\rho) = \0{T}{2 L^3}\sum_{n_\mu\in\Z} G_k(q,\rho)\,\dot R(q)\,, 
\quad {\rm with}\quad q_\mu = 2 \pi n_\mu \0{1}{L_\mu}\,.  
\end{align}
Note that the effects of finite volume are qualitatively similar in
other dimensions $(d-1) + 1$. Thus, for the purpose of this work, we
restrict ourselves to $(3+1)$ dimensions. In the following, we discuss
results obtained in the $\phi^4$-approximation for flows at both
finite temperature and finite volume. These results are obtained by
integrating the flows (containing loop integrations \eq{eq:loopfinV}),
\eq{eq:LPAobserve}, over the RG scale $k$ from $\Lambda$ to 0.

In the current FRG setup the dimensional reduction discussed in the
last \sec{sec:range} is apparent from the flow. To begin with,
at finite temperature and $k/T\ll 1$ the flow only receives
contributions from the Matsubara zero mode (for regulators that decay
sufficiently fast for frequencies $q_0/T \ll 1$). Consequently, in this
regime the flow reads
\begin{align}\label{eq:flowT0mode}
  \dot V_{ k\ll T }(\rho)\to \0T2 \int_{\vec q} \0{\dot
    R_k(q_0=0,|\vec q|)}{ \vec q^2+R_k(q_0=0,|\vec q|)+m_k(\rho)}\,.
\end{align}
The prefactor $T$ can be absorbed with an appropriate rescaling of
$V_k, \rho$, leading to dimensionally reduced flows for   
\begin{align}\label{eq:thermalres}
V_{\text{\tiny{3d}}}= V/T\,,\quad \rho_{\text{\tiny{3d}}}=  \rho/T\,. 
\end{align}
The new variables \eq{eq:thermalres} have the momentum dimensions of
the 3-dimensional potential and field-squared. Moreover,
\eq{eq:flowT0mode} takes the form and the dimensional properties of a
(3+0)-dimensional flow for the potential.

With the same line of arguments it can be easily seen that a finite
volume setup does not support a non-vanishing condensate, effectively
reducing to quantum mechanics for a large correlation length: if the
cutoff scale $k$ is far smaller than the inverse size of the system,
$k\ll 1/L$, only the spatial zero mode contributes to the flow,
effectively reducing it to a $0+1$-dimensional flow with only the
frequency integral to be performed. This mimics a quantum mechanical
flows, to wit
\begin{align}\label{eq:flowL0mode}
  \dot V_{ k\ll 1/L }(\rho)\to \0{1}{2 L^3} \int \0{\dm q_0}{2
    \pi}\0{\dot R_k(q_0,|\vec q|=0)}{q_0^2+ R_k(q_0,|\vec
    q|=0)+m^2_k(\rho)}\,.
\end{align}
As in \eq{eq:flowT0mode} the prefactor $1/L^3$ can be absorbed with a
similar rescaling of $V_k$ and $\rho$ similarly to \eq{eq:thermalres},
with
\begin{align}\label{eq:Vres}
  V_{\text{\tiny{1d}}} = V \,L^3 \,,\quad \rho_{\text{\tiny{1d}}}=
  \rho\, L^3 \,. 
\end{align}
This renders \eq{eq:flowL0mode} in a form that is identical to that of
the quantum-mechanical (0+1)-dimensional flow for the potential. The
latter flow does not admit a non-trivial minimum for the potential:
there is no spontaneous symmetry breaking in quantum mechanics.

Finally, at finite temperature and for $k\ll \min(1/L, T)$, the system
reduces to the zero mode,
\begin{align}
\label{eq:0dim} 
\dot V_{ k\ll  \min(1/L, T) }(\rho)\to \0{T}{L^3} \0{k^2}{k^2+m^2_k(\rho)}\,, 
\end{align}
where, without loss of generality, we have used that $R_k(0) = k^2$ and
$\dot R_k(0) = 2 k^2$. Again this can be mapped into the standard flow
by a rescaling of $V_k$ and $\rho$ similarly to
\eq{eq:thermalres},\eq{eq:Vres} with
\begin{align}\label{eq:VTres}
  V_{\text{\tiny{0d}}} = V\,L^3/T \,,\quad
  \rho_{\text{\tiny{0d}}}= \rho\, L^3/T\,,
\end{align}
i.e.\ effectively taking $T/L^3\to 1$. This is the flow of a
(0+0)-dimensional field theory, the generating functional being the
one dimensional integral
\begin{align}\label{eq:0+0Z}
  Z_{\text{\tiny{0d}},k}(J)= \int d \phi\, \exp \left\{ -
    V_{\text{\tiny{0d}},\Lambda}(\rho) -k^2 \rho_\text{\tiny{0d}} +
    J\, \phi_\text{\tiny{0d}}\right\} \,,
\end{align}
where $\Lambda \ll \min(1/L, T)$. This trivial flow has been studied
in e.g.\ Ref.~\cite{Keitel:2011pn}: it does not admit a non-trivial
minimum for the respective effective action
$\Gamma_{\text{\tiny{0d}}}[\phi]$.

\subsection{Condensates}\label{sec:cond}
In a finite volume no long range spatial fluctuations are
permitted. However, the latter ones are required for phases with
spontaneous symmetry breaking and non-vanishing condensates. Only in
one or more spatial directions and with infinite extent (at vanishing
temperature) non-vanishing condensates survive for the present Ising
universality class model ($O(1)$-symmetry).

The disappearance of condensates in a finite volume is easily seen in
$O(N)$-models for $N>1$ due to the presence of massless Goldstone
modes with mass functions $m^2_{k,\theta}(\rho)=V_k'(\rho)$, where
$\vec \theta$ are the $N-1$ Goldstone modes. Here, emphasising the
differences to the Ising-type case $N=1$, we briefly discuss this
case.

The flow of the minimum can be derived from the $t$-derivative of the
equation of motion, $V'_k(\rho_0)=0$ for $\rho_0>0$.  Its flow reads
\begin{align}\label{eq:flowrho0}
  \partial_t  V_k'(\rho_0) = \dot V'(\rho_0) +\dot \rho_0
  V''(\rho_0)=0\,,
\end{align}
where the flow $\dot V_k'$ is given in \eq{eq:dtVp}. The masses of the
Goldstone modes vanish in the broken phase, i.e.\ at a non-trivial
minimum $\rho_0>0$. By resolving \eq{eq:flowrho0} for $\dot\rho_0$ we
arrive at
\begin{align}\nonumber 
  \dot \rho_0 =& -\0{\dot V'(\rho_0)}{V''(\rho_0)} \\[2ex]\nonumber =
  &\012 \int_q \dot R_k(q)\left[\left(3+\rho_0\0{\lambda_3}{\lambda_2}
    \right) G_{k,\rho}^2(q)+
    (N-1) G_{k,\theta}^2(q)\right]\\[2ex]
  \to &\0{1}{k^2} \0{T}{L^3}\left[
    \0{3+\rho_0\0{\lambda_3}{\lambda_2}}{\left(1+\0{2
          \rho_0\lambda_2}{k^2}\right)^2 } + (N-1)\right]\,.
  \label{eq:min} 
\end{align} 
The first equality in \eq{eq:min} provides the flow of $\rho_0$ in the
local potential approximation in an $O(N)$-model. The first term in
the second line is the radial contribution while the second term comes
from the Goldstone modes with the massless propagator $G_\theta$. In
$d\leq 2$ (infinite range) dimensions and $N>1$ the latter term does not
tend to zero in the limit $k\to 0$ while the first term does with
powers of $k^2/m_k^2$. The Goldstone term is positive, and hence the
condensate is driven to zero with $k\to 0$, that is
$t\to-\infty$. This reflects the Mermin-Wagner theorem.

In a finite volume and at finite temperature the flow reduces to the
third line for $k\ll \min(T,1/L)$. Again, the singularity in the flow
is apparent. As a consequence, $\rho_0$ is driven to zero with $k\to 0$. This is also seen in
Yukawa-type models such as the linear quark-meson models and
atom-condensate models for ultracold atoms, see e.g.\
\cite{Braun:2005gy,Braun:2010vd,Braun:2011uq,Braun:2011iz,Tripolt:2013zfa}.

Note that this argument does not apply to $N=1$ with the discrete
$Z_2$-symmetry. The $O(1)$ model (Ising universality class) has no
Goldstone mode and the vanishing of the condensate for $d<2$
dimensions cannot be read-off directly from the flow of the
condensate, \eq{eq:min}. In particular, the flow in \eq{eq:min}
reduces at finite temperature and volume to
\begin{align}
  \dot \rho_0 \to k^2\0{T}{L^3}
  \0{3+\rho_0\0{\lambda_3}{\lambda_2}}{(k^2+2 \rho_0\lambda_2)^2}\,,
  \label{eq:minN1} 
\end{align}
for $k\ll \min( T, 1/L)$. This relates to the fact that the Ising
universality class, $N=1$, shows a phase transition in $d=2$. Below two
dimensions, $d<2$, it does not show a phase transition. 

We remark that this structure is not seen in a Taylor expansion about
the flowing minimum $\rho_0$. For example, in the
$\phi^4$-approximation the condensate does not necessarily vanish at
vanishing cutoff scale: for a finite mass $m_k^2$ the flow of all
couplings, including $\rho_0$ and the free energy $F_k$, can
vanish with powers of $k^2/m_k^2$. Indeed, the non-vanishing minimum
 stays at higher orders at this expansion too. This hints at a failure
of the polynomial expansion about $\rho_0$. Note, however, that an
expansion about $\rho=0$ works in dimensions $d\leq 2$, and converges
rather rapidly with the full solution of the flow equation for the
potential. This suggests an expansion about the flowing minimum
$\rho_0$ for $k\gtrsim 1/L$ and one about $\rho=0$ for $k\ll 1/L$.

Still, the vanishing of the condensate at finite volume and
temperature is easily seen. For the sake of completeness we briefly
present the related argument. Assume for the moment that
$m_{k=0}^2(\rho_0)> 0$. Then, for $k>0$ the potential has a non-convex
regime with $m_k^2(\rho<\rho_w)< 0$ and the turning point $\rho_w$
with $m_k^2(\rho_w)=0$. For this regime the relative strength of the
free energy $\Delta V_k= V_k(\rho=0) -F_k$ is governed by
\begin{align}\label{eq:DeltaF_k}
  \Delta \dot V_k= \012 \int_q
  \dot{R}_k(q)\,\left( G_k(\rho, q) -G_k( q)\right)\,,
\end{align} 
where $\rho\leq \rho_w$ is a $k$-independent field value in the
non-convex regime with $m_k^2(\rho<\rho_w)<0$, potentially leading to
large flows close to the singularity triggered by $k^2 +
m_k^2(\rho)=0$.  For the current purpose it is sufficient to use the
estimate 
\begin{align}%
\0{1}{p^2+R_k(p^2)  }< \0{1}{p^2+R_k(p^2) + m_k^2(\rho<\rho_w)}
\end{align}%
 and, hence, we estimate
\begin{align}\nonumber 
  \Delta \dot V_k >&\, \012 \int_q \dot{R}_k(q)\,\left( \0{1}{p^2
      +R_k} - \0{1}{p^2 +R_k
      +m_k^2}\right)\\[2ex]
  = &\,\012 \int_q \dot{R}_k(q)\,\left( \0{1}{p^2 +R_k} m_k^2
    \0{1}{p^2 +R_k+m_k^2}\right)\,.
\label{eq:DeltaF_kfin}\end{align}
In the limit $k/[\min(1/L,T)]\to 0$ we arrive at 
\begin{align}
  \Delta \dot V_k \to \0{T}{L^3}\,.
\label{eq:DeltaF_kfinIR}\end{align} 
Evidently, the flow of $\Delta V_k$ does not vanish for $k\to 0$ and
hence the difference turns negative: the minimum is at $\rho=0$.
Note that an iteration of this argument also excludes the existence of
several minima and, therefore, guarantees a smooth vanishing of the minimum
$\rho_0$ with $k\to 0$.

For vanishing temperature or more generally infinite range dimensions
$d<2$ the above argument can be generalised by discussion of the flow of the
turning point $\rho_w$ of the radial mass function $m_k^2(\rho)$ with
$m_k^2(\rho_w)=0$.  We have $\rho_w\leq \rho_0$ for the $O(1)$-model
and $\rho_w = \rho_0$ in the $O(N>1)$-model. The full analysis is beyond the
scope of the present work and hence discussed elsewhere.

\section{Thermodynamics in a finite volume }\label{sec:thermodyn}

One of the main goals of this work is the study of finite volume
effects in the pressure, 
\begin{align}\label{eq:defp}
p = -\partial F/\partial {\CV}\,,
\end{align} 
with the free energy $F$ and the spatial volume $\CV=L^3$. The free energy $F$
relates to the effective action $\Gamma[\phi_{\text{\tiny EoM}}]$ on
the solution of the EoM, $\phi_{\text{\tiny EoM}}$. It
can be normalised (shifted) such that it vanishes in the combined
infinite volume and zero temperature limit, that is
\begin{align}\label{eq:Fnorm}
  F = T(\Gamma^{T,L}[\phi_{\text{\tiny EoM}}] -
  \Gamma^{0,\infty}[\phi_{\text{\tiny EoM}}]) \,.
\end{align}
\Eq{eq:Fnorm} is the free energy related to thermal and finite volume
fluctuations, and the pressure \eq{eq:defp} with \eq{eq:Fnorm} is the
combined finite volume and thermal pressure. In the following we use
the free energy density $f_k$ already introduced in \eq{eq:observablesf}
in the cutoff scale-dependent case. We have
\begin{align}\label{eq:f}
  f_k=\0{F_k}{\CV}=\0{\Gamma_k[\phi_{\text{\tiny EoM}}]}{\beta\CV}\,, \qquad
  {\rm with} \quad \beta=\0{1}{T}\,,
\end{align}
by extracting the trivial spatial volume factor in the static free
energy.  \Eq{eq:Fnorm} and \eq{eq:f} entail that the pressure $p_k$ 
comprises the fluctuations induced from finite temperature and
volume. The difference of free energies densities $f_k$ at finite and
infinite volume relates to the Casimir force, for a FRG computation
 see Ref.~\cite{Jakubczyk:2012iza}.

The flow of the pressure is then given by the one of the free energy
density and its volume or, equivalently, length derivative,
\begin{align}\label{eq:flowpV}
  \dot p_k(T) = - \dot f_k(T,L)-\0L3 \0{\partial \dot
  f_k(T,L)}{\partial L}\,,
\end{align}
where we have used that $\CV=L^3$. The flow of the free energy
density $f_k$ is given by
\begin{align}%
\dot{f}_k(T,L) = \012 \sum_q G^{{T},L}_k(q) \dot{R}_k(q) -
 \012  \int_q G^{0,\infty}_k(q) \dot{R}_k(q)\,, 
\label{eq:flowf}
\end{align}%
where the sign $\sum_q$ stands for a sum over frequencies and discrete
spatial momenta. In \fig{fig:dflengths} the flow of the free energy
density, \eq{eq:flowf}, is shown for different volumina.

\bfc%
\includegraphics[width=\columnwidth]{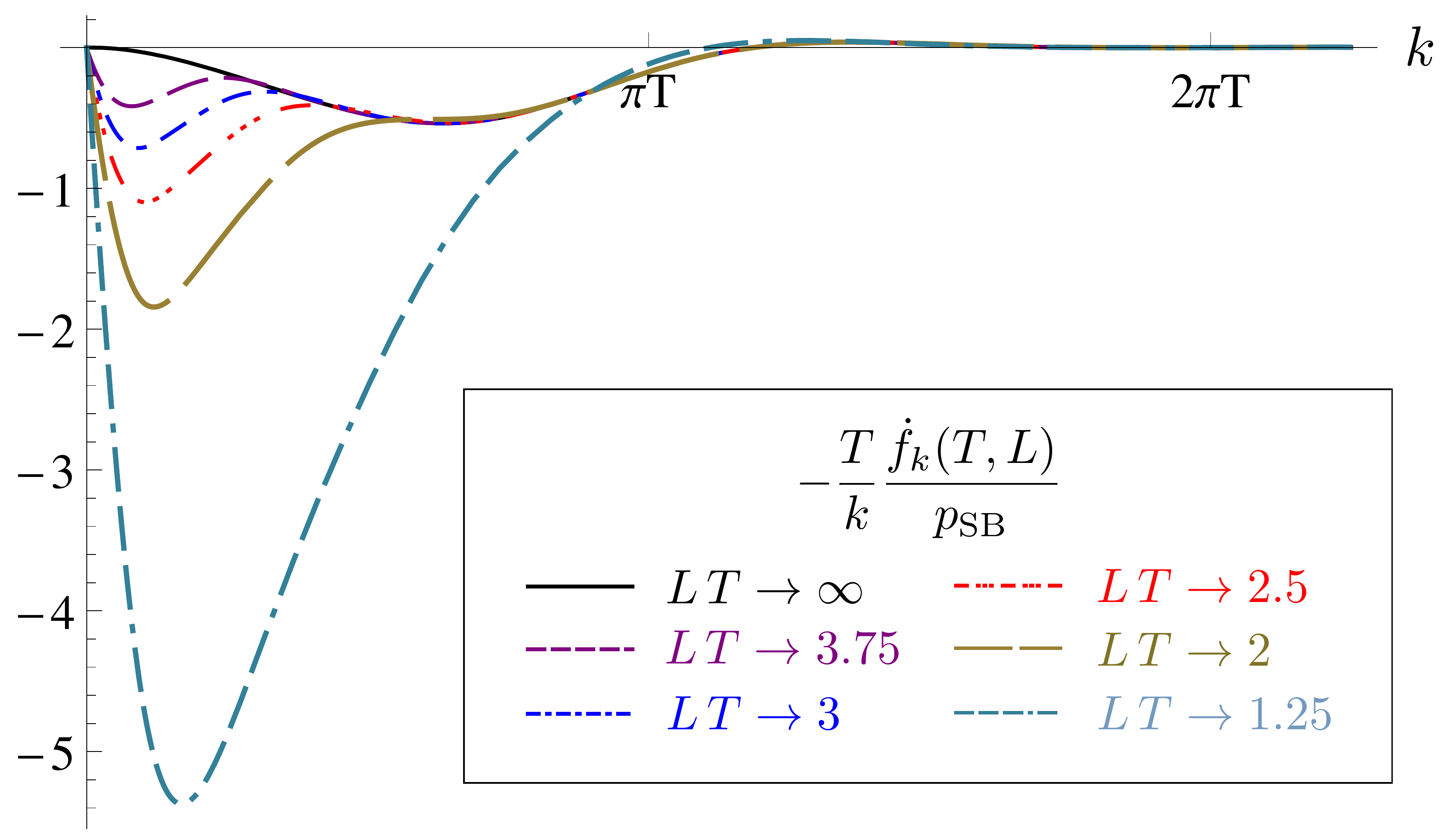}
\caption{Spatial volume dependence of the flow of the free energy
  density $f_k$ for a coupling $\lambda_\Lambda=0.5$ and an exponential
  regulator with $m=1$, cf. eq.~\eq{eq:Rexp}.}
\label{fig:dflengths}
\efc%
In \eq{eq:flowf} the first term is the finite temperature flow
involving the thermal and/or finite volume propagator
$G^{T,L}_k(q)$, whereas the second term is the normalisation in
terms of the vacuum propagator $G^{T=0,L=\infty}_k(q)$. Moreover, in
the present LPA approximation the flow of the free energy density
$\dot f_k$ at a given cutoff scale $k$ is simply given by the
difference of the flows of the effective potential,
\begin{align}\label{eq:flowpLPA}
\dot f_k =  \dot V^{T,L}_k(\rho_0)-\dot V^{0,\infty}_k\,,
\end{align}
cf. \sec{sec:FRG}. The flow equation for the infinite
volume pressure is given by that of the free energy density,
\begin{align}\label{eq:flowp}
\dot p_k(T) = - \lim_{L\to \infty} \dot f_k(T,L)\,,
\end{align}
as the $L$-derivative of $\dot f_k$ vanishes faster than $1/L$. The
flow of the thermal pressure $\dot p_k$, \eq{eq:flowp}, is shown for
different regulators in \fig{fig:dpSBT}.
\bfc%
\includegraphics[width=\columnwidth]{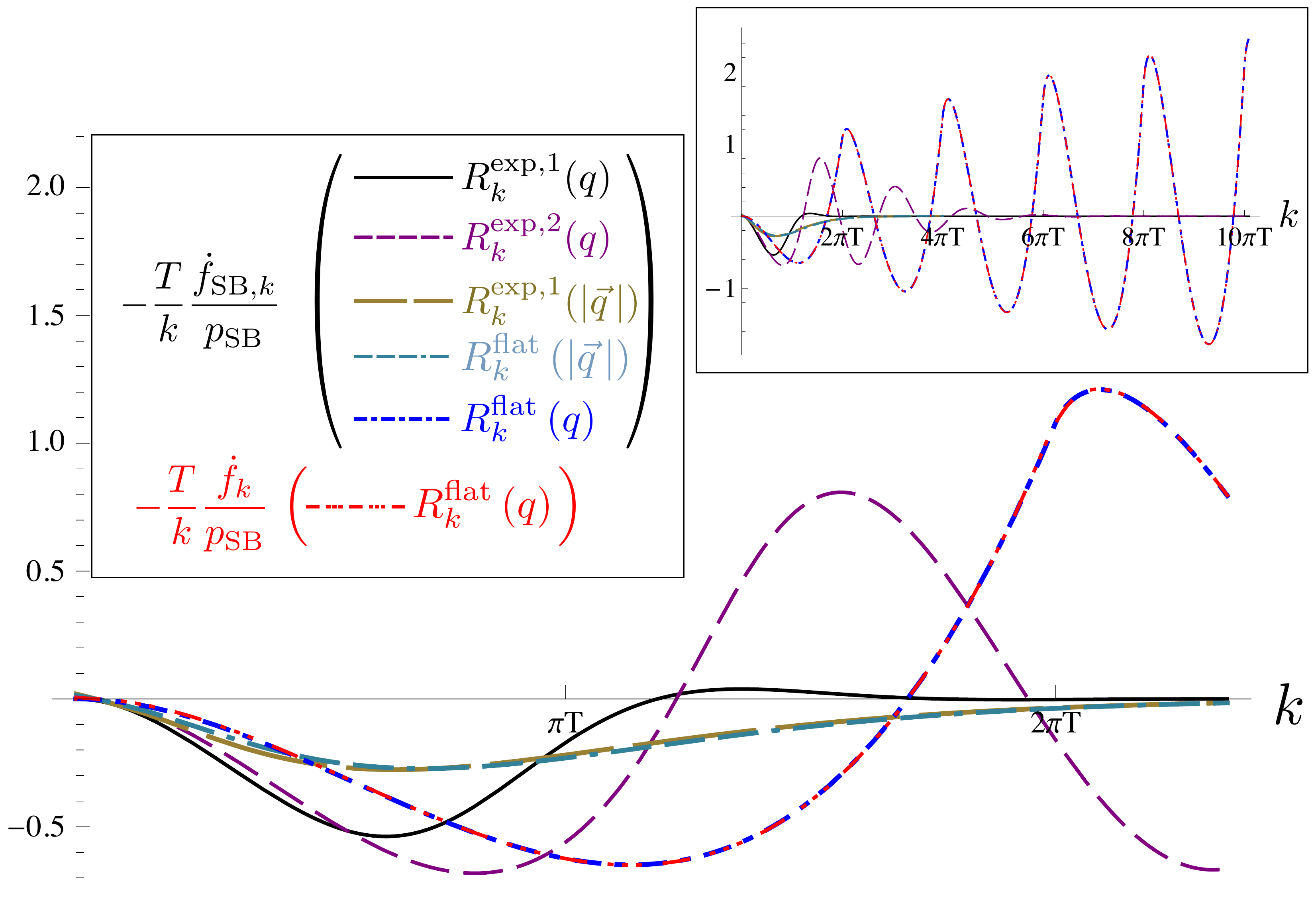}
\caption{Flow of the Stefan--Boltzmann pressure
  $p_{\text{\tiny{SB}},k}=-f_{\text{\tiny{SB}},k}(T,\infty)$ for
  different regulator, and the full pressure for
  $R_k^{\text{\tiny{flat}}}$, in an infinite volume, with $\lambda_\Lambda=0.5$ for the full result.}
\label{fig:dpSBT}
\efc%

The influence of the second term in a finite volume can readily be
discussed qualitatively. It is already well-known from purely thermal
flows that $\dot f_k>0 $ for $k\lesssim \min( T, 1/L)$. It is this
cutoff regime where the main contribution to the pressure is collected
during the flow, leading to a positive pressure, see also
\fig{fig:dflengths}.  For larger cutoff scales the flow switched sign,
$\dot f_k \lesssim 0$ for $k\lesssim \min( T, 1/L)$.  As the volume and
temperature pressure grows with decreasing volume and/or increasing
temperature the above properties imply
\begin{align}\nonumber 
  \0{\partial\dot f_k}{\partial L} & <0\quad {\rm for}\quad k\lesssim
  \min( T, 1/L)\,,\\[2ex]
  \0{\partial\dot f_k}{\partial L} & \gtrsim 0\quad {\rm for}\quad
  k\gtrsim \min( T, 1/L)\,.
\label{eq:Vderf}\end{align}
In summary the integrated contributions from $-1/3\, L \partial_L \dot
f_k$ decrease the pressure in a finite volume, and the free energy
density $f_{k=0}$ gives an upper estimate for the pressure.

\subsection{Stefan--Boltzmann pressure for three-dimensional flat
  regulator}\label{sec:bSB3d}

A simple but instructive example for the properties and definitions
discussed above is the tree-level free energy density
$f_{\text{\tiny{SB}},k}$ and pressure $p_{\text{\tiny{SB}},k}$, i.e.\ the
Stefan--Boltzmann contributions, for two reasons.  Firstly, it
illustrates on an analytic level the computations and results
presented in the next Section~\ref{sec:results}.  Secondly, and more
importantly, it demonstrates constraints on the regulators emerging
from finite temperature or volume in a clear way.

First, we compute $p_{\text{\tiny{SB}}}$ in the infinite volume limit
with the three-dimensional flat regulator, see \eq{eq:Rflat}. This
regulator is often used as analytic computations can be pushed
far. The free propagator in the presence of
$R^{\text{\tiny{flat}}}_k(\left| \vec q\right | )$ is simply given by
\begin{align}%
  G^{(0)}_k(q)= \0{1}{q_0^2+k^2}\theta(k^2-\vec q^2)+\0{1}{q_0^2+\vec q^2}
  \theta(\vec q^2-k^2)\,.
\label{eq:propopt}\end{align}%
With the propagator in \eq{eq:propopt} and $\dot
R^{\text{\tiny{flat}}}_k= 2 k^2 \theta(k^2-\vec q^2)$, the flow for the
infinite volume Stefan--Boltzmann pressure \eq{eq:flowp} can be given
analytically, see e.g.\ \cite{Litim:1998nf,Litim:2006ag},
\begin{align}\nonumber 
  \dot{p}_{\text{\tiny{SB}},k}(T) =& -\int_0^{k}\0{\dm
    q_s}{2\pi^2}\left(T\!\!\!  \sum_{n=-\infty}^{\infty} \frac{k^2
      q_s^2}{k^2+\omega_n^2} -
    \int_{q_0}  \frac{k^2 q_s^2 }{k^2+q_0^2}\right)\\[2ex]\nonumber 
  = &-\frac{k^4}{12 \pi ^2} \left(\coth\frac{k}{2 T}-1\right)
\end{align}  
For large RG scales compared to the temperature this turns into
\begin{align}
  \stackrel{k/T\to\infty} {\longrightarrow} &-\frac{k^4}{6 \pi ^2}\,
  \exp \left(-\0{k}{T} \right) \,, 
\label{eq:SB_flow}\end{align}%
where $q_s = \sqrt{\vec q^2}$ and $\int_{q_0} =
\int_{-\infty}^{\infty}\0{\dm q_0}{2\pi}$. The flow \eq{eq:SB_flow}
and that for other regulators is shown in
\fig{fig:dpSBT}. \Eq{eq:SB_flow} entails that for large cutoff scales
$k/T\to\infty$ the flow decays exponentially with $\exp(-k/T)$ in line
with \eq{eq:Tdecay}. This is expected as the cutoff used in
\eq{eq:SB_flow} does not affect the frequency sum, and reflects the
fast decay of thermal fluctuations in the presence of large mass
scales. Hence, the pressure at the initial scale $\Lambda\gg T$ can be
safely put to zero, $p_{\text{\tiny{SB}},\Lambda}=0$. The integration
over the RG scale $k$ from $\infty$ to $0$ gives the correct pressure
$p_{\text{\tiny{SB}}}$ of a gas of free scalar particles,
\begin{align}
  p_{\text{\tiny{SB}}}(T)= \int_{\infty}^{0}\frac{\dm k}{k}\
  \dot{p}_{\text{\tiny{SB}},k} (T)\ = \ \0{\pi^2 T^4}{90}\,.
\end{align}%
The Stefan--Boltzmann expression \eq{eq:SB_flow} is easily upgraded to
the full flow.  In the latter case we simply have to substitute $k^2
\to k^2 +m_k^2$ in the denominators in the first line of
\eq{eq:SB_flow} and arrive at 
\begin{align}\nonumber 
  \dot{p}_k(T) =& -\frac{k^4}{12 \pi ^2} \0{1}{\sqrt{
        1+\0{{m^{T}_k}^2}{k^2} }}\left( 
    \coth\0{k}{2T}\sqrt{1+\0{{m^{T}_k}^2}{2}}-1\right)\\[2ex] 
& -\frac{k^4}{12 \pi ^2}  \left( \0{1}{\sqrt{
        1+\0{{m^{T}_k}^2}{k^2} }}- \0{1}{\sqrt{
        1+\0{{m^{0}_k}^2}{k^2} }} \right)\,.
\label{eq:pflow3d}\end{align}%
The term in the first line of \eq{eq:pflow3d} is decaying
exponentially for large cutoff scales $k/T\to\infty$ also in the
presence of non-vanishing mass function $m_k^T$. In turn, the term in
the second line is proportional to $\left(m_k^T\right)^2-\left(m_k^{T=0}\right)^2$: the
exponential decay of the full flow in \eq{eq:pflow3d} hinges on the
exponential decay of the difference of the mass functions at finite
temperature and vanishing temperature. The flow in \eq{eq:pflow3d}
already encodes the one of differences of the effective potential with
$m_k\to m_k(\rho)$. Evidently, the exponential decay for
$\left(m_k^T(\rho)\right)^2-\left(m_k^{T=0}(\rho)\right)^2$
is encoded in the flow. This simple
but instructive example already illustrates the main features of the
thermal and finite volume properties under investigation: the flow
reflects the thermal decay with large mass scales in case that this is
also present for the flowing couplings.

We now repeat this computation with the three-dimensional flat regulator 
in the general case with a finite volume and at finite
temperature. For the sake of simplicity we first drop the $\partial_L
\dot f_k$-term in \eq{eq:flowpV}. The three-dimensional spatial integral
turns into sums, and we arrive at the Stefan--Boltzmann expression for
the flow of the free energy density,
\begin{align}\nonumber 
  \dot{f}_{\text{\tiny{SB}},k} =& \0{k}{2 L^3} \sum_{\vec q^2\leq k^2}
  \coth\frac{k}{2 T} -
  \int_{q_0}  \frac{k^2 q_s^2 }{k^2+q_0^2}\\[2ex]
  = & \0{k}{2 L^3}\sum_{\vec q^2\leq k^2} \left(
    \coth\frac{k}{2 T}-1\right) - \left( \frac{k^4}{12 \pi ^2}-\0{k}{2
      L^3 }\sum_{\vec q^2\leq k^2}\right)\,,
  \label{eq:SB_flow3dfV}\end{align}%
in a splitting similar to that in \eq{eq:pflow3d}. In
\eq{eq:SB_flow3dfV} we have
\begin{align}\label{eq:qs2}
\vec q^2= \left(\0{2\pi}{L}\right)^2 \vec n^2\,,\quad {\rm with}\quad 
\vec n=(n_1,n_2,n_3)\,.
\end{align}
Results for different regulators are shown in \fig{fig:dfSBV}. 
\bfc%
\includegraphics[width=\columnwidth]{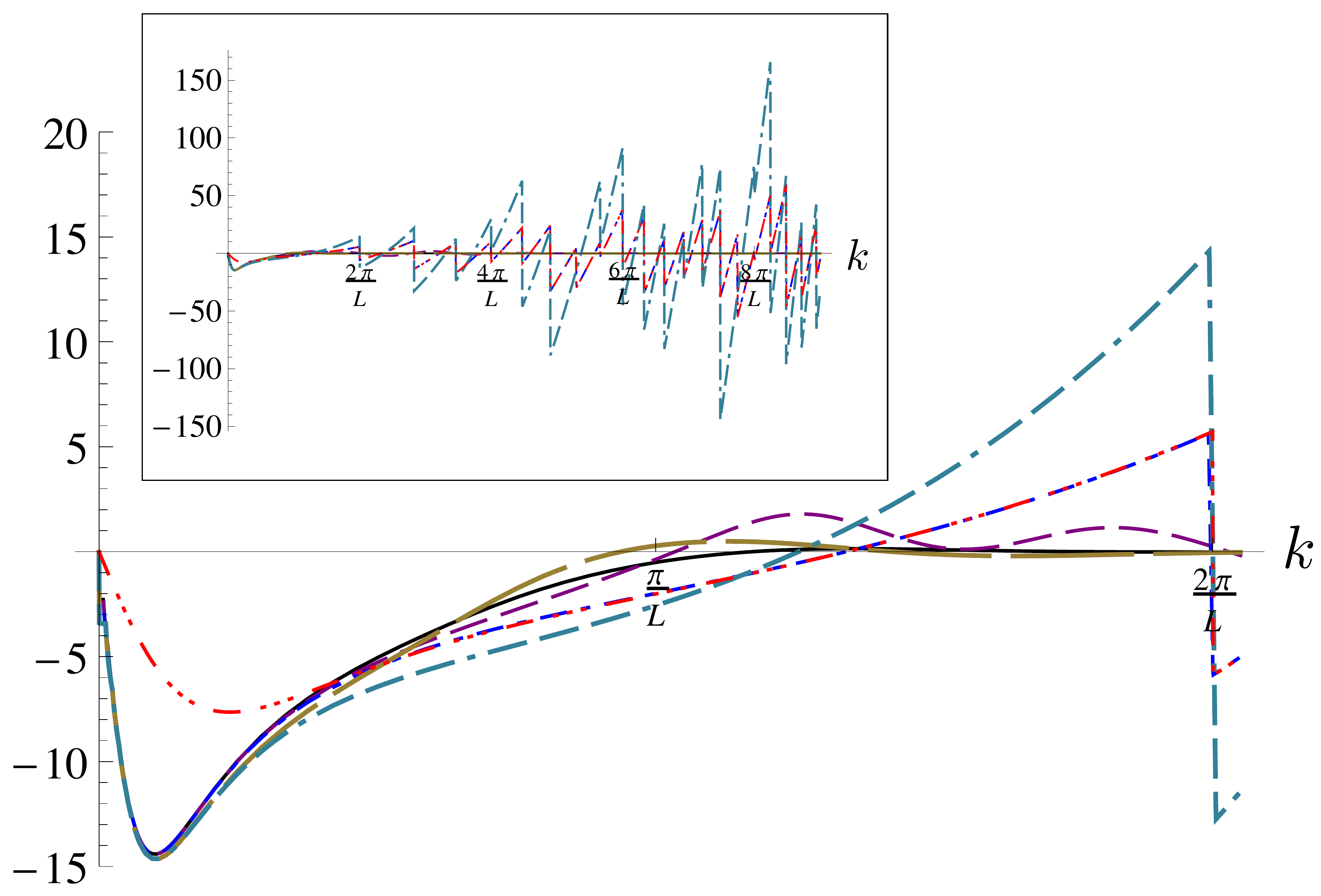}
\caption{Flow of the Stefan--Boltzmann free energy density
  $-f_{\text{\tiny{SB}},k} $ for different regulators, and and the
  full free energy density, $-f_{k} $, for $R_k^{\text{\tiny{flat}}}$,
  in an finite volume with relative lengths $L\,T=1$ and
  $\lambda_\Lambda=0.5$ for the full result. The colour coding and
  normalisation is the same as in \fig{fig:dpSBT}.}
\label{fig:dfSBV}
\efc%

The first term in parenthesis in the second line of
\eq{eq:SB_flow3dfV} decays exponentially with $\exp(-k/T)$ for large
$k/T$, but does not decay exponentially with $k\, L$ for fixed $k/T$.
This signals the failure of the three-dimensional flat regulator to
reflect the exponential decay of finite volume fluctuations in the
presence of large mass scales in a similar decay with the cutoff
scale. 

For a direct access to the finite volume decay we take the limit $T\to
0$. Then, the first term in parenthesis vanishes and we are left with
the second term. This limit gives the flow of the free energy density
in a finite volume normalised with that in an infinite volume. It is
expected to decay exponentially with $k\, L\to\infty$ similar to a
thermal flow. Instead, the combined flow fluctuates about zero with a
rising enveloping amplitude proportional to $k^2$.

The sum in the second term simply counts the number of momentum points
in a three-dimensional sphere with radius $k$. Within the numerical
precision of the present work we find for a $d$-dimensional sphere,  
\begin{align}\label{eq:numbers}
  \sum_{q^2\leq k^2} \to 
  \0{\pi^{d/2}}{ \Gamma\left(\0{d+2}{2}\right)} \left(\0{k L}{2 \pi}\right)^d
+O(k^{\theta_d})\,,\qquad \theta_{d\geq 3} =d-2\,,
\end{align}
where $q^2=(2 \pi)^2/L^2(n_1^2 +\cdots +n_d^2)$ is the $d$-dimensional
discrete momentum squared. For $d=1$ we have trivially
$\theta_1=0$. Note that while the sub-leading behaviour in
\eq{eq:numbers} has been proven for $d\geq 4$, it is still subject to
ongoing research for $d=2,3$, with $d=2$ being the Gau\ss\ circle
problem. The best estimates we found in the literature are
$\theta_3\leq 21/16$ \cite{HeathBrown} and $1/2 <\theta_2 \leq 131/208$ 
\cite{Huxley}. In Appendix~\ref{app:flatbox}, for illustration,
we have performed a similar computation for a regulator that achieves
a flat propagator in a spatial momentum box with box lengths
$2k$. Such a regulator allows for fully analytic computations, and is
even more adapted to the symmetries of the spatial box.

With \eq{eq:numbers} the Stefan--Boltzmann flow $\dot
f_{\text{\tiny{SB}},k}$ in a finite volume diverges proportional to
$k^2$ for the three-dimensional flat regulator. This asks for a
non-vanishing initial condition $f_{\text{\tiny{SB}},\Lambda}$ at the
initial cutoff scale $k=\Lambda$ in order to guarantee that
$f_{\text{\tiny{SB}},k=0}$ is the tree level free energy density.
Corrections of the thermodynamics due to the full dynamics are encoded
in the fluctuating part
\begin{align}\nonumber 
  f_k-f_{\text{\tiny{SB}},k} \propto & -\frac{k^4}{12 \pi ^2}\left(
    \0{1}{\sqrt{ 1+\0{{m^{0,\infty}_k}^2}{k^2} }}- \0{1}{\sqrt{
        1+\0{{m^{T,L}_k}^2}{k^2} }}\right)\\[2ex]  & +\left[\0{k}{2
      L^3 }\sum_{\vec q^2\leq k^2} -\frac{k^4}{12 \pi ^2}\right]\left(
    \0{1}{\sqrt{ 1+\0{{m^{T,L}_k}^2}{k^2} }}-1 \right)\,,
\label{eq:p+Delp}\end{align} 
where we have dropped the exponentially decaying parts. The term on
the right hand side of the first line in \eq{eq:p+Delp} is
proportional to $k^2\left((m^{T,L}_k)^2-(m^{0,\infty}_k)^2\right)$. The mass
difference tends to a constant, albeit small (up to logarithmic
corrections), for large scales. In turn, the second line is
proportional to the difference of three-dimensional integral and
three-dimensional sum, that is the sub-leading term in
\eq{eq:numbers}.  Hence, its total scaling is proportional to
$k^{1+\theta_3}\approx k^2$. 
\bfc%
\includegraphics[width=\columnwidth]{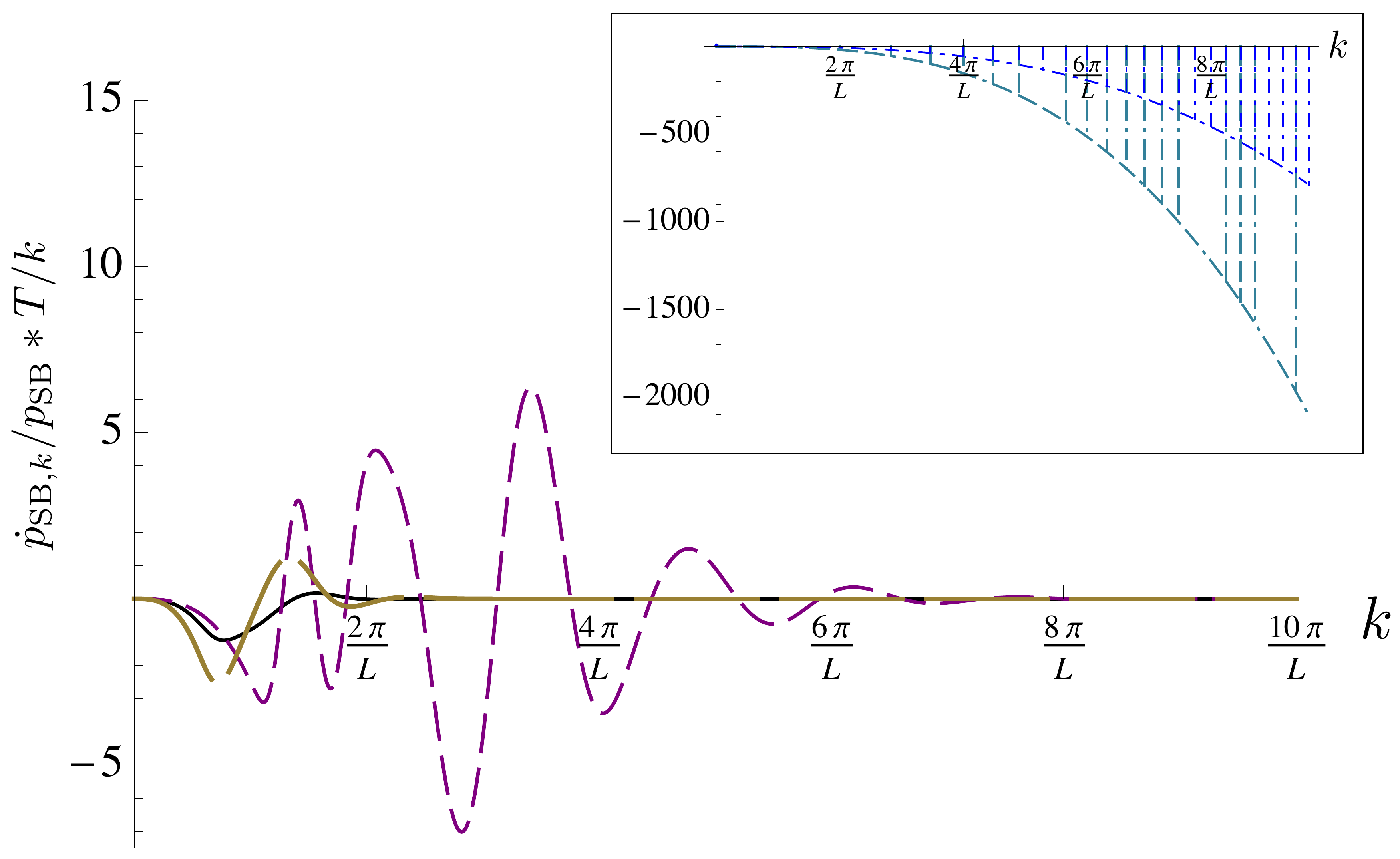}
\caption{Flow of the pressure $p_{\text{\tiny{SB}},k} $ for different
  regulators in a finite volume with relative lengths $L\,T=1$. The
  colour coding and normalisation is the same as in \fig{fig:dpSBT}.}
\label{fig:dpSBV}
\efc%

The above analysis also entails that for the present regulator,
$R_k^{\text{\tiny{flat}}}(\vec q^2)$ the second term in the flow of
the pressure, \eq{eq:flowpV}, a priori cannot be dropped even in the
infinite volume limit. It can be easily computed from
\eq{eq:SB_flow3dfV}, and reads
\begin{align}
  -\0L3 \0{\dot{f}_{\text{\tiny{SB}},k} }{\partial L}=& \dot
  f_{\text{\tiny{SB}},k} +\frac{k^4}{12 \pi ^2} - \013 \0{k^3}{L^3}
  \coth\frac{k}{2 T}\,\sum_{\vec n\in \Z^3}\delta(k^2-\vec q^2)\,,
\label{eq:VdVf3d}\end{align}
where the second term is vanishing everywhere except for $k^2 = (2
\pi/L)^2 (n_1^2+n_2^3+n_3^3)$, simply collecting the contributions at
the discrete momentum values. Using both terms, \eq{eq:SB_flow3dfV}
and \eq{eq:VdVf3d}, in the flow equation for the pressure,
\eq{eq:flowpV}, we arrive at the simple expression
\begin{align}
  \dot p_{\text{\tiny{SB}},k} =& \frac{k^4}{12 \pi ^2} -\013
  \0{k^3}{L^3} \coth\frac{k}{2 T}\,\sum_{\vec n\in
    \Z^3}\delta(k^2-\vec q^2) \,.
\label{eq:pSBflow3d}\end{align}
The flow of the Stefan--Boltzmann pressure for the three-dimensional
flat regulator is given by a $k^4$-term and a $\delta$-function
contribution that peaks at the discrete spatial momenta. For the full
pressure flow we can repeat the same analysis as for the flow of the
free energy density leading to similar $k^2$-corrections.  

In \fig{fig:dpSBV} we show the flow of the free energy density for the
three-dimensional flat regulator in comparison to their
Stefan--Boltzmann counterparts. Again, the differences are negligible
for large cutoff scales as $m_k^2/k^2 \ll 1$. This limit is usually
achieved in theories or models with small ultraviolet fluctuations in
the validity range of the theories. Note that in the present
$\phi^4$-theory the Landau pole is hit at a large cutoff
$k_{\text{\tiny{Landau}}}$, and one has to safely stay below this
scale. We also emphasise that both, the flow of the free energy
density, $-\dot f_k$ in \eq{eq:SB_flow3dfV}, and that of the pressure,
$\dot p_k$ in \eq{eq:pSBflow3d}, tend towards \eq{eq:SB_flow} in the
infinite volume limit. For $T\to 0$ and $L\to \infty$ both flows
vanish.

The simple example of thermodynamical observables studied in the
present Section has already taught us an important lesson: flows with
non-analytic regulators do not reflect the physical exponential decay
with large mass scales in the decay with the cutoff scale. While the
latter property is not necessary in order to guarantee the former
physical one, it is potentially cumbersome to achieve in
approximations.

\subsection{Stefan--Boltzmann pressure for four-dimensional flat
  regulator}\label{sec:bSB4d}

For the four-dimensional flat regulator, \eq{eq:Rflat}, the tree-level
propagator reads
\begin{align}%
  G^{(0)}_k(q)= \0{1}{k^2}\theta(k^2-q^2)+\0{1}{q^2}
  \theta(q^2-k^2)\,, 
\label{eq:propopt4}\end{align}%
with the momentum squared $q^2$ only taking discrete values 
\begin{align}\label{eq:q2}
 q^2 = (2 \pi T)^2 n_0^2+ \left(\0{2 \pi}{L}\right)^2\vec
  n^2\,, \quad n_\mu\in\Z\,,
\end{align}
and $\vec n=(n_1,n_2,n_3)$. We have chosen all lengths to be equal,
$L_i=L$. With the propagator \eq{eq:propopt4} and 
$\dot R_k^{\text{\tiny{flat}}}(q)=2 k^2 \theta(k^2-q^2)$, the
flow of the free energy density is given by
\begin{align}
  \dot{f}_{\text{\tiny{SB}},k}=& \0{T}{L^3} \sum_{q^2\leq k^2}  -
    \int_{q^2\leq k^2}\0{\dm^4 q}{(2 \pi)^4}= 
\0{T}{L^3} \sum_{q^2\leq k^2}     -    \frac{k^4}{32 \pi ^2} \,.
  \label{eq:SB_flowfV}\end{align}%
Similarily to the computation above for the three-dimensional
flat regulator, we have to add the second term in
\eq{eq:flowpV}. This leads to the simple flow
\begin{align}
  \dot p_{\text{\tiny{SB}},k} =&\frac{k^4}{32 \pi ^2} \ -\023
  \0{T}{L^3}\sum_{n_\mu\in \Z} \vec q^2 \delta(k^2-q^2) \,,
\label{eq:pSBflow4d}\end{align}
see \fig{fig:dpSBV}. The spatial and four-dimensional momenta squared
$\vec q^2$ and $q^2$ are defined in \eq{eq:qs2} and \eq{eq:q2},
respectively.

Similar to the case of the three-dimensional flat regulator we
conclude that the flow of the free action density is sensitive to
finite volume effects for all cutoff scales for the four-dimensional
flat regulator. With a similar analysis as in the last
Section~\ref{sec:bSB3d} it follows that there are subleading terms
proportional to $k^2\left( (m^{T,L}_k)^2-(m^{0,\infty}_k)^2\right)$
and $k^2 (m^{T,L}_k)^2$.  For the choice $m_k^2/k^2 \ll 1$, being the
initial value in both finite and infinite volume as well as zero and
non-zero temperature, at large cutoff scales the corrections are
small.  In \fig{fig:dfSBV} we show the flow of the free energy density
for the four-dimensional flat regulator in comparison to their
Stefan--Boltzmann counterparts, as well as the Stefan--Boltzmann flows
for other regulators. 
\bfc%
\includegraphics[width=\columnwidth]{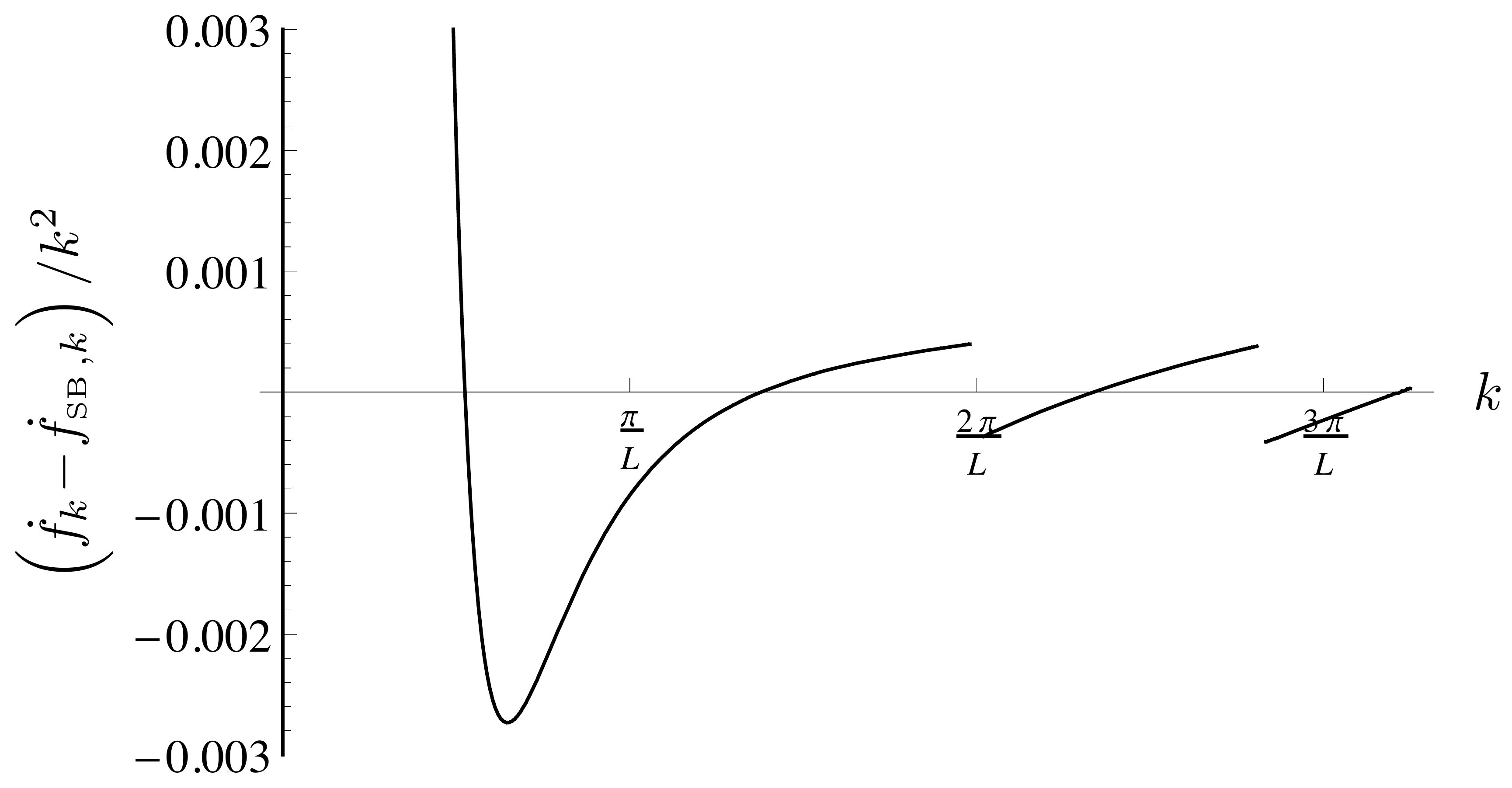}
\caption{The difference of the flow of the free energy density
  contribution for the four-dimensional flat regulator
  $R_k^{\text{\tiny{flat}}}(q)$.}
\label{fig:Deltafdecay}
\efc%
In \Fig{fig:Deltafdecay} we show the difference
of the flow of the free energy density. This difference clearly shows
the ultraviolet sensitivity of the flow with respect to finite volume effects.

In the infinite volume limit both, \eq{eq:SB_flowfV} and
\eq{eq:pSBflow4d} tend towards the infinite volume flow. For $T\to 0$
and $L\to \infty$ both flows vanish. We also conclude that for both,
the three-dimensional and four-dimensional flat regulators, the
thermodynamics is approximately accessible for $m_\Lambda^2/\Lambda^2
\ll 1$ at the expense of a trivial non-vanishing initial condition
$f_\Lambda=f_{\text{\tiny{SB}},\Lambda}$ and
$p_\Lambda=p_{\text{\tiny{SB}},\Lambda}$. However, since the flow of
$f_k$ and $p_k$ only depends on the couplings, the Stefan--Boltzmann
terms and the $\Delta \dot f_k$, $\Delta \dot p_k$ can be integrated
separately: the total free energy density/pressure is simply given by
\begin{align}\label{eq:totalintfp} 
  p_{k=0}\approx p_{\text{\tiny{SB}}}(T,L)+\int_\Lambda^0 \0{dk}{k}
  \Delta \dot p_k\,,
\end{align}
with the trivial Stefan--Boltzmann pressure $p_{\text{\tiny{SB}}}(T,L)$
in a finite volume and the integrated flow of $\Delta \dot p_k$.

The findings of the present and the last section carry over to the
theory in the presence of three- and four-dimensional sharp
regulators, \eq{eq:Rsharp}. Indeed, in these cases the ultraviolet
growth is very similar. Below \eq{eq:Rsharp} it has been discussed
that the theory in the presence of sharp regulators is directly
related to respective ultraviolet momentum cutoff regularisations of
DSEs in a finite volume and at finite temperature,
\cite{Fischer:2007pf,Luecker:2009bs,Bonnet:2011hh,Bonnet:2012az}.
Consequently the respective correlation functions miss the exponential
decays unless one applies a volume and/or temperature dependent
renormalisation procedure. Note that only for the present
thermodynamical observables, pressure and free energy density, this
shows up as a growth with the ultraviolet cutoff scale $\Lambda$, in
higher order correlation functions the missing exponential decay is
hidden in polynomially decaying terms. 

We add that these findings
apply to all situations with a sharp momentum cut-off for discrete
momentum modes. This includes in particular Landau level sums in the
presence of magnetic fields. There it is well-known that the critical
temperature, and therefore the phenomenon of (inverse) magnetic
catalysis shows an exponential sensitivity with respect to the scales of the
problem. Hence it is particularly sensitive to the presence or
absence of the exponential decay discussed above. For FRG-applications
to the (inverse) magnetic catalysis see
\cite{Skokov:2011ib,Fukushima:2012xw,Andersen:2014oaa,%
  Braun:2014fua,Mueller:2015fka}, for other approaches with
ultraviolet momentum cutoffs including DSEs and perturbation theory,
see the recent reviews \cite{Andersen:2014xxa,Miransky:2015ava}.

We close with the remark that it is not simply the missing
differentiability of the regulator that spoils the decay with the
cutoff scale. It originates in the non-analytic behaviour of the
regulator at $p^2 =k^2$. In Appendix~\ref{app:css} we test the smooth
modifications of the flat regulator suggested in \cite{Nandori:2012tc}
that are $\CC^\infty$ at $p^2 =k^2$ at the expense of an essential
singularity $\propto \exp\{ -{\rm const.}/(1-p^2/k^2)\}$ for spatial
momentum approaching $k$ from below, $p^2\to k^2_-$. They show the
same missing exponential decay in the flow, although the smoothening
of the non-analyticity leads to smaller fluctuations at large cutoff
scales, see \fig{fig:css} in Appendix~\ref{app:css}. This is not
surprising as it is well-known from the thermal case that the
exponential decay reflects the pole structure of the loop integrals in
the complex plane if the Matsubara sum is rewritten as the (original)
contour integral. This suggests that analytic versions of the flat
regulator have all the necessary properties for optimising flows at
finite volume and temperature in the sense of the optimisation
criterion in \cite{Pawlowski:2005xe}.

\subsection{Pressure for general
  regulators}\label{eq:bSB3d}

The flow of the pressure for general regulators is based on that of
the free energy density and its length derivative. More generally the
length-derivative of the flow of the effective action is given by
\begin{align}\nonumber 
  L\0{\partial \dot \Gamma_k}{\partial L} =& \sumint_q \vec
  q^2 \partial_{\vec q^2} \left[ G_k\,\dot R_k(q)\right] 
\\[2ex]  &
-\012 \sumint_q
  G_k\,\dot R_k\,G_k(q) L\left. \0{\partial 
  \Gamma_k^{(2)}(q)}{\partial L}\right|_{\vec q^2}\,,
\label{eq:dLG}\end{align} 
where the last term hits the $L$-dependences of the
vertices. \Eq{eq:dLG} can be inserted in the definition of $f_k$,
\eq{eq:f}, to obtain the $L$-derivative of the general flow of the
free energy density. In the present LPA-approximation the second term
in the flow of the pressure, \eq{eq:flowpV}, reduces to
\begin{align} 
  - \0{L}{3}\0{\partial \dot f_k}{\partial L} =& \dot f_k +\dot
  V_k^{0,\infty}+\016 \sumint_q \left( 2 \vec q^2 \partial_{\vec q^2}
    - L \0{\partial m_k^2}{\partial L} \partial_{m_k^2}\right)
  G_k\,\dot R_k\,,
\label{eq:derVfLPA}
\end{align} 
see \fig{fig:dpSBV}. This leaves us with the task of computing the
$L$-dependence of the couplings. This is either done by performing
computations at $L$ and $L+\epsilon$ and taking discrete derivatives
or by using flow equations for the $L$-derivatives in a fixed volume,
see Appendix~\ref{app:derV}.

We close the present Section with a brief summary of the results.  In
the Figs.~\ref{fig:dpSBT}, \ref{fig:dfSBV}, \ref{fig:dpSBV} the
findings of this Section are illustrated. In \fig{fig:dpSBT} the
Stefan--Boltzmann pressure in an infinite volume is shown for different
regulators and the full pressure for the four-dimensional flat
regulator. One sees both the exponential decay with $k$ for analytic
regulators (in the frequency $p_0$) and the power law rise for
non-analytic regulators (in the frequency). All flows are apparently
smooth in contradistinction to the finite volume flows depicted in
Figs.~\ref{fig:dfSBV}, \ref{fig:dpSBV}. This is related to the fact
that the chosen regulators with a $q_0$-dependence are functions of
$q_0^2+\vec q^2$. This entails that the jumps for non-analytic
regulators, that are related to the discrete Matsubara frequencies are
smoothened out by the spatial momentum integration. Regulators of the
form $R_k(q_0,\vec q)= R_{0,k}(q_0^2)+R_{s,k}(\vec q^2)$ with
non-analytic regulators $R_{0,k}$ lead to jumps similarly to that
shown in \fig{fig:dfSBV}.

In \fig{fig:dpSBV} the Stefan--Boltzmann pressure in a
finite volume is shown for different regulators. Again one sees both
the exponential decay with $k$ for analytic regulators and the
polynomial rise for non-analytic regulators. Additionally, one has
discontinuities as described in \eq{eq:SB_flow3dfV}, \eq{eq:SB_flowfV}
for the flow of the free energy density and in \eq{eq:VdVf3d},
\eq{eq:pSBflow4d} for the flow of the pressure. The polynomial rise of
the free energy density and the pressure with $k^3$ and $k^4$ in a
finite volume for the three-dimensional and four-dimensional flat
regulator, respectively, hint at similar, but reduced problems for the
flow of the mass $m_k^2$ (or $\kappa_k$) in both cases. The $k$-power
counting of these flows has an additional factor $k^{-2}$ naively
leading to rising volume-dependence with $k$ and $k^2$,
respectively. If the naive power counting holds, some care is required
with the initial conditions for the mass, as then it varies with the
volume and the temperature.

For the coupling constant $\lambda_k$ we expect a $1/k$ suppression of
the flow for the three-dimensional flat regulator and a
$k^0$-dependence for the four-dimensional flat regulator. This
translates into a slow convergence of the flow towards the infinite
volume flow in the first case, hence asking for a very large initial
cutoff scale.  In the latter case this leads to a logarithmic
dependence with $k$ of the initial condition on the volume and
temperature, and requires some care.

\section{Couplings in a finite volume}\label{sec:results}

We have already discussed in the introduction and in more details in
Section~\ref{sec:finite}, that a quantitative grip on the thermal,
volume and cutoff scale dependence is pivotal for getting hold of in
particular thermodynamical quantities and the equation of
state. Moreover, the analytic results in the last
Section~\ref{sec:thermodyn} indicate the problems of non-analytic
regulator choices such as the flat regulator and the sharp cutoff.

\subsection{Thermal \& finite volume decay}\label{sec:TVdecay}
In the following, we discuss the quantitative behaviour of the flows
of couplings and thermodynamical observables in dependence of the
choice of regulators. Despite the choice of regulator both finite
temperature and volume are pure infrared effects, their effects are
damped above energy scales which are typically of the order of the
first frequency either in temporal or spatial direction.  In other
words, for scales much larger than the volume or temperature the
theory turns into the infinite volume vacuum theory. The details of
the decay are physically important, for the decay behaviour in mass
scales and momentum scales see \eq{eq:Tdecay}, \eq{eq:Tpoldecay} and
\eq{eq:TpoldecayG} for the important example of the Tan-contact.

\bfc%
\includegraphics[width=\columnwidth]{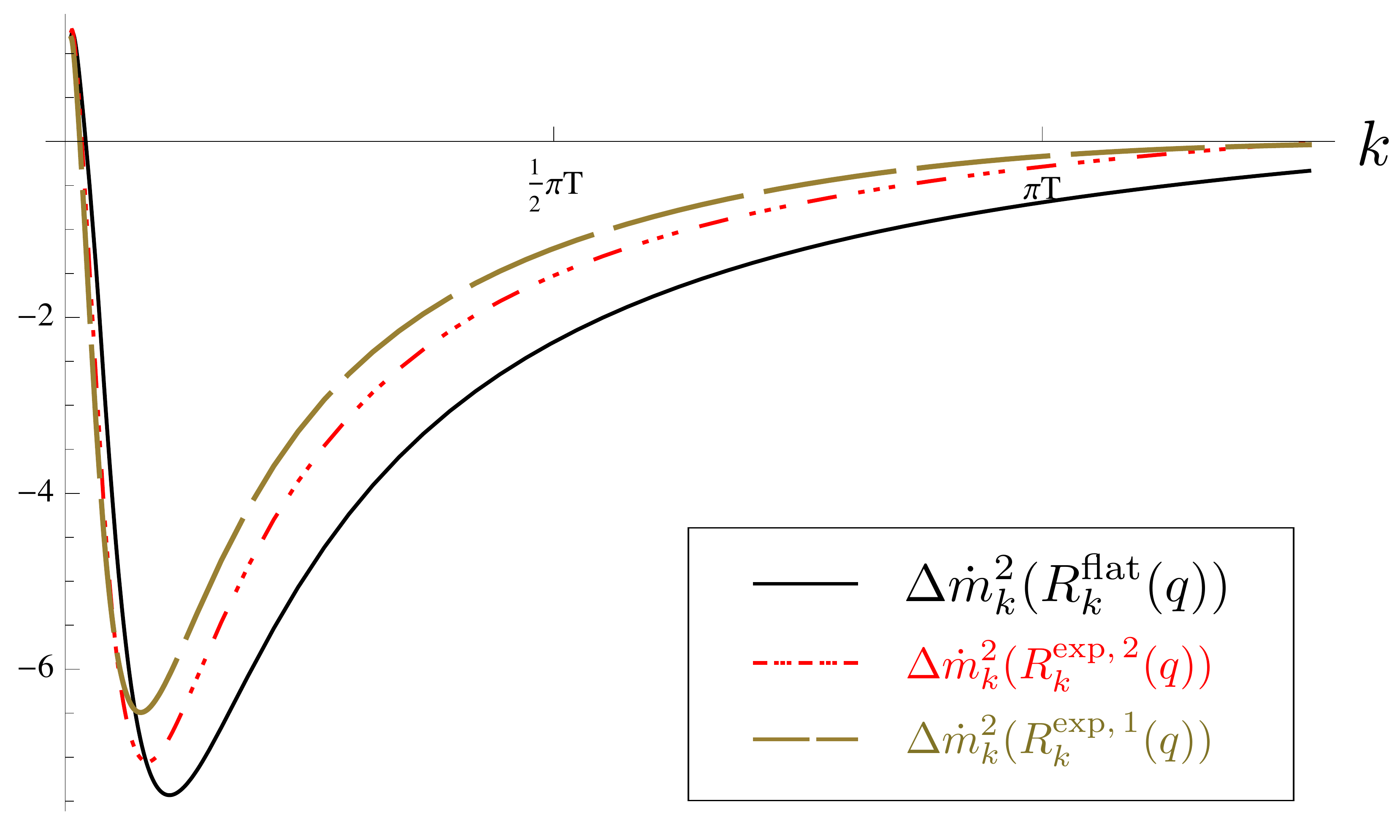}
\caption{Normalised mass difference $\Delta m_k^2$ at finite
  temperature $T=0.25$ in an infinite volume as defined in \eq{eq:DeltaO}.}
\label{fig:thermal_mass}
\efc%
\bfc%
\includegraphics[width=\columnwidth]{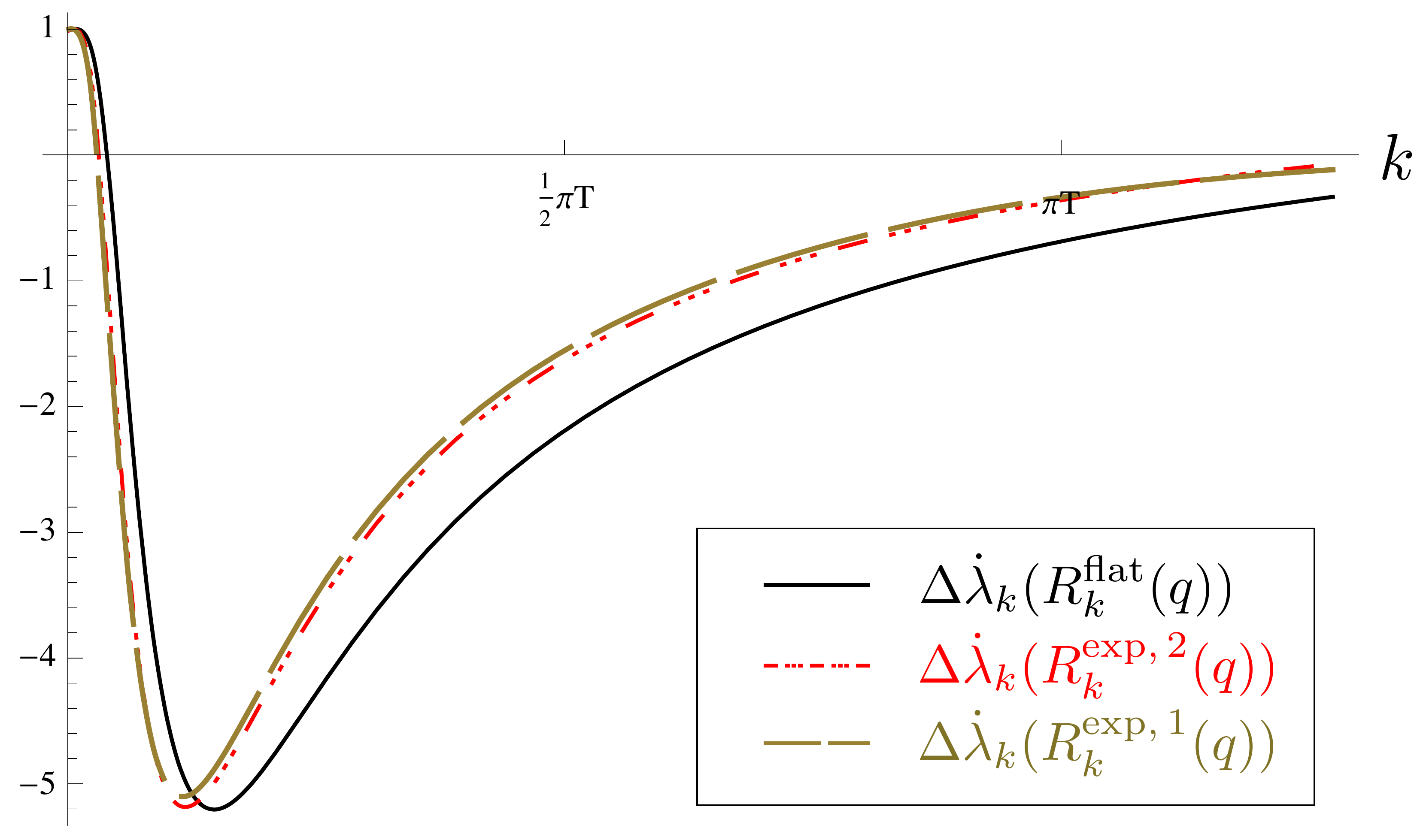}
\caption{Normalised $\phi^4$-coupling difference $\Delta\lambda$ at
  finite temperature $T=0.25$ in an infinite volume as defined in
  \eq{eq:DeltaO}.}
\label{fig:thermal_coupling}
\efc%
In methods where fluctuations at all scales are integrated out at once
this usually happens around the typical frequency or spatial momentum
scale given by the first frequency $2\pi T$ of $2\pi /L$. This feature
does not necessarily hold for the momentum cutoff scale $k$ flow. As
has been already seen in Section~\ref{sec:thermodyn} for
thermodynamical observables, this cutoff scale depends strongly on the
shape of the regulator. More precisely, it depends on the way the
infrared modes are suppressed: the sharper this suppression is
implemented, the further in $k$ one has to go to lose sensitivity
to thermal or finite volume effects. In particular for non-analytic
regulators such as the flat regulator and the sharp cutoff this loss
of sensitivity is delayed to infinity.

For the regulator shapes given in \eq{eq:Rexp}, \eq{eq:Rflat} and
\eq{eq:Rsharp} regulating either four-dimensionally or in spatial
directions only, the thermal flow of the tree-level pressure
$\dot{p}_{\text{\tiny{SB}},k}$ is given in \fig{fig:dpSBT}. In an
infinite volume the problem of a "mode-counting" cutoff can be
circumvented by regulating the spatial momenta only. In contrast, in a
finite volume this generic property rules out sharp and flat
regulators for practical computations. We note that this is not cured
by the inclusion of running masses and couplings, as also seen in
\fig{fig:dfSBV} anticipating results below.

We conclude that the exponential thermal decay in \eq{eq:Tdecay} does
not simply hold with $m_{\rm gap}\to k$, even though this has been
observed for the three-dimensional flat regulator, see
\eq{eq:SB_flow}. In general, one has to allow for a prefactor $c_T$
with
\begin{align}\label{eq:mgapk}
  \exp\left(- \0{m_{\rm gap}}{T}\right) \to \exp\left(- c_T
    \0{k}{T}\right) \,,\quad \quad c_T\in[0,1]\,.
\end{align}
For regulators that only depend on spatial momenta and have an
(effective) infrared mass $R_k(0)=k^2$ the upper limit, $c_T=1$, is
saturated: the sum over the Matsubara modes is not cut-off, and the
thermal behaviour is untouched.  With the dominant mass scale $k$ for
$k\to\infty$ this leads to the standard thermal suppression with
$c_T=1$. Note that strictly speaking the upper limit is not well
defined as it depends on the mass scale introduced with the
regulator. Only if the latter scale is identical with $k$ the
prefactor $c_T$ satisfies \eq{eq:mgapk}.

In turn, for regulators that are non-analytic in frequency space such
as the four-dimensional flat regulator, the exponential suppression is
lost and the lower bound is saturated, $c_T=0$.

In a finite volume the decay towards the vacuum limit is induced by
both, thermal and volume fluctuations. Hence, it leads to a
generalisation of the thermal decay \eq{eq:Tdecay}, \eq{eq:mgapk}. For
\eq{eq:flowf}, we parametrise the decay in the limit
\begin{align}
  \lim_{k/T,k\,L\to \infty } \left| \Delta {\mathcal{O}}_k^{T,L}
  \right| \propto A\left(\0kT,\, k\,L\right) {\rm exp}\left( -
    c_T(L\,T)\0{k}T \right)\,,
\label{eq:kdecay}\end{align}%
where the function $A$ now encodes thermal and volume effects, and
\begin{align}\label{eq:DeltaO}
  \Delta {\mathcal{O}}_k^{T,L} \equiv \0{{\mathcal{O}}_k^{T,L}-
    {\mathcal{O}}_k^{0,\infty}}{{\mathcal{O}}_k^{0,\infty}}\,.
\end{align}
$\Delta {\mathcal{O}}_k^{T,L}$ is the difference of an observable
${\mathcal{O}}_k^{T,L}$ at finite volume and temperature and infinite
volume and vanishing temperature, normalised by its flow at
$T=1/L=0$. In \fig{fig:thermal_mass} and \fig{fig:thermal_coupling}
the normalised difference $\Delta\CO_k$ is shown for the mass and the
coupling difference respectively for finite temperature in an infinite
volume.

The definition \eq{eq:DeltaO} also applies to flows or diagrammatic
kernels of flows, the threshold functions. Flow equations of
thermodynamical observables and vertices depend on the cutoff
derivative of the regulator and powers of propagators. The decay of
general flows can hence be measured by that of
\begin{align}\label{eq:COn}
\CO^{T,L}_{k,n}=\int_q G_k(q)^n\, \dot R_k(q)\,. 
\end{align}
In \fig{fig:dRG1234} we show the decay of general flows for different
regulators, precisely, we show
\beq%
\0{\CO^{T,L}_{k,n}-\CO^{0,\infty}_{k,n}}{k^{d-2\left(n-1\right)}}\,,
\label{eq:dRG1234}
\eeq%
with the definition \eq{eq:COn} for $n=1,2,3,4$. We normalise with
respect to the canonical dimension $k^{d-2\left(n-1\right)}$.  In the
presence of an exponential decay, that is $c_{T}\neq 0$ for $\Delta
\CO^{T,L}_{k,1}$, all $\Delta \CO^{T,L}_{k,n}$ show the same decay as
they can be generated by $n-1$ derivatives with respect to $m_k^2$ from
$\Delta \CO^{T,L}_{k,1}$.

\begin{figure}[t]\begin{center}%
\includegraphics[width=\columnwidth]{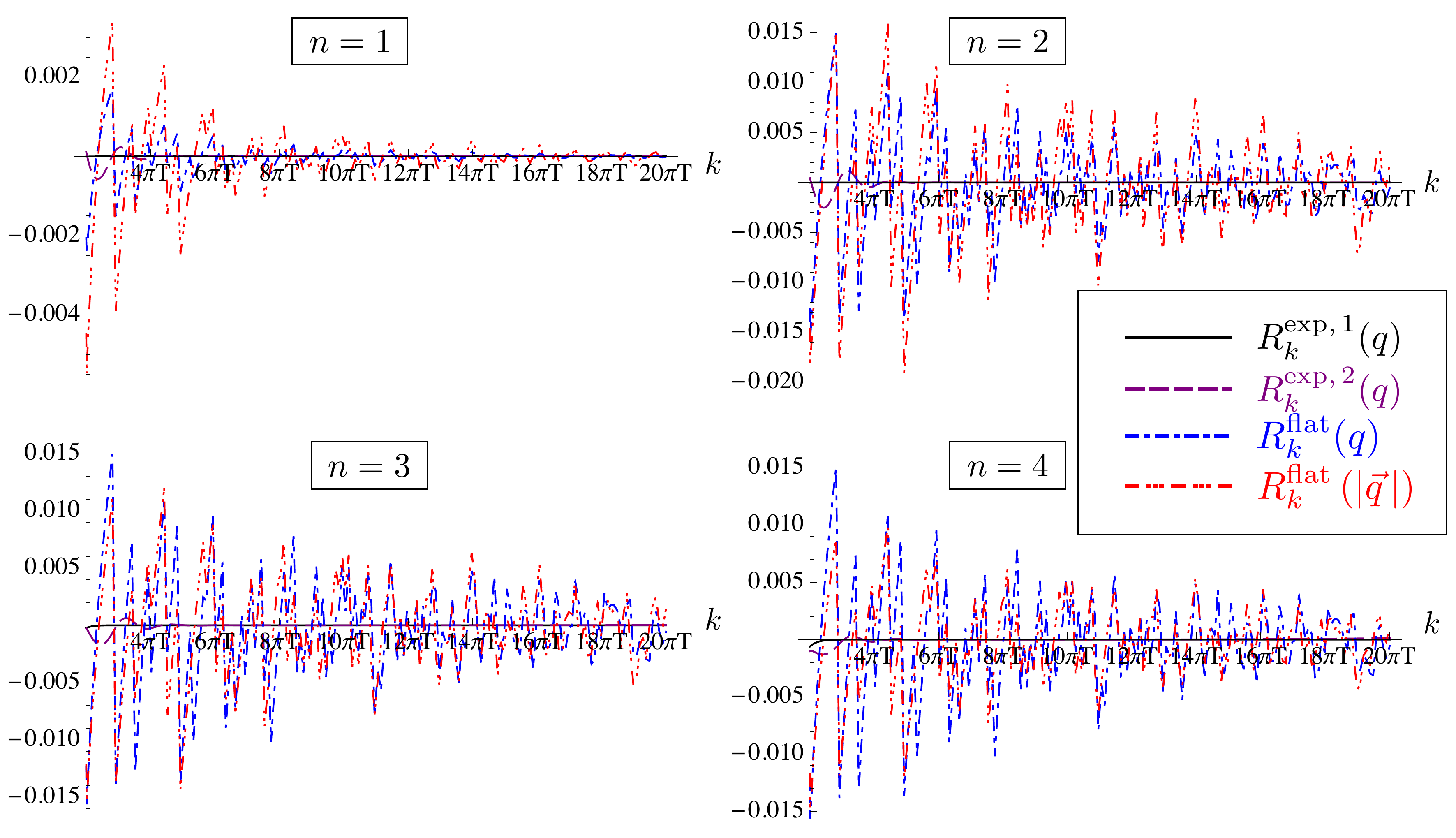}
\caption{Flows given by \eq{eq:dRG1234} for $n=1,2,3,4$ in a finite
  volume for different regulators.}
\label{fig:dRG1234}
\efc%

The thermal decay at infinite volume and the volume decay at vanishing
temperatures are given by the respective limits. For $L\,T\to \infty$,
\eq{eq:kdecay} gives the thermal decay \eq{eq:mgapk} with
$c_T=c_T(\infty)$, that is
\begin{align}
  \lim_{k/T\to \infty} \left| \Delta
      {\mathcal{O}}_k^{T,\infty}\right| \propto
  A\left(\0kT,\, \infty\right) {\rm exp}\left( - c_T\0{k}T \right)\,,
\label{eq:kTdecay}\end{align}%
see \fig{fig:thermal_mass}, \fig{fig:thermal_coupling}. This is the
analogue of \eq{eq:Tdecay} with the substitution \eq{eq:mgapk}.  In
turn, for $L\,T\to 0$, \eq{eq:kdecay} describes the volume decay at
vanishing temperature, and we have
\begin{align}\label{eq:c_L}
\lim_{L\,T\to 0} \0{c_T(L\, T)}{L\,T} = c_L \,,
\end{align}  
and the equivalent relation to the purely thermal decay,
\eq{eq:kTdecay} for the pure finite volume case,
\begin{align}
  \lim_{L k\to\infty } \left|\Delta {\mathcal{O}}_k(0,L)\right| \propto
  A\left(\infty,\, k\,L\right) {\rm exp}\left( - c_L\,k\,L \right)\,. 
\label{eq:kLdecay}\end{align}%
The function $c_T(L\,T)$ interpolates between the asymptotic thermal
scaling and the asymptotic finite volume scaling, and depends on the
shape of the regulator, $c_T(L\,T)=c_T(R_k;L\,T)$. It can be extracted
in the limit of $k\rightarrow \infty$ by
\begin{align}%
\label{eq:cTV}
c_{T}(L\,T) =- \0Tk \lim_{k\rightarrow \infty} \log{\left\{ \left|
      \Delta{\mathcal{O}}_k(T,L) \right| \right\}}\,. 
 \end{align}%

 In \fig{fig:cTV} we study the approach of the right hand side of
 \eq{eq:cTV} towards $c_T$ for $n=1,2,3,4$ and an exponential
 regulator, \eq{eq:Rexp}, with $m=2$, in both the infinite and finite
 volume case, in the latter case we use $L=1/T$. The local minima
 indicate the local maxima and minima in the positive and negative
 envelopes of the decay, as seen in \fig{fig:dpSBT} and
 \fig{fig:dfSBV} for $n=1$. We find that the asymptotic limits for
 $k/T$ and/or $k/\left(1/L\right)$ are reached quickly above the
 characteristic scale induced by $c_T$.

\bfc%
\includegraphics[width=\columnwidth]{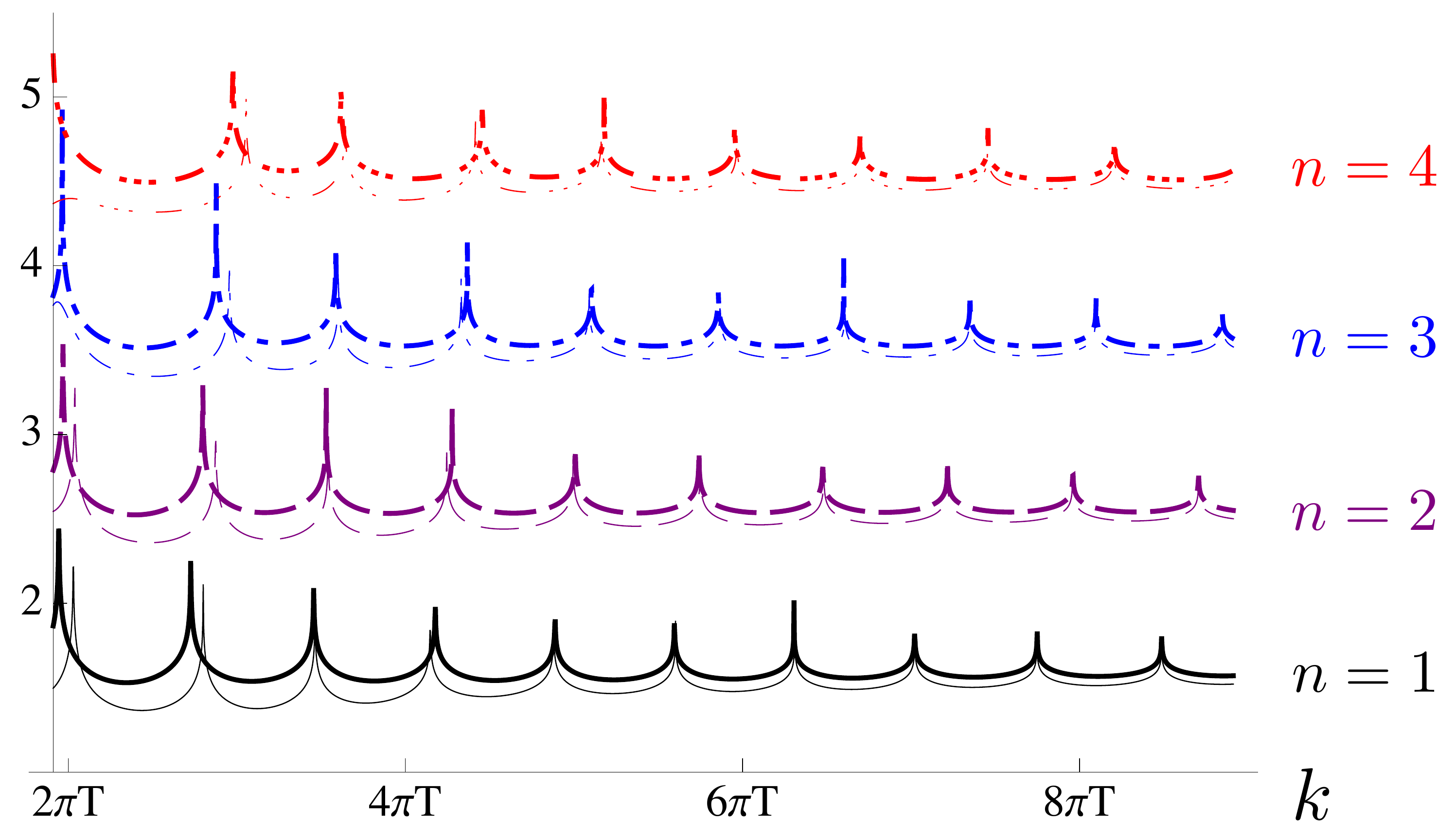}
\caption{Approach of flows towards the vacuum limit for scales $k$
  larger the temperature and volume as parametrised by \eq{eq:cTV} and
  the an exponential regulator, \eq{eq:Rexp}, with $m = 2$. Thick (thin,
  lower) curves are in infinite (finite with $L=1/T$) volume. }
\label{fig:cTV}
\efc%
 In \fig{fig:cTVinterpolate} we study the interpolation of $c_{T}$
 between the asymptotic forms at $T=0$ and $L=\infty$ for the
 exponential regulators with $n=1,2$. For this figure we have
 extracted $c_T$ with the value of the minimum at maximal $k$, in
 e.g.\ \fig{fig:cTV} the minimum between $8\pi T$ and $10 \pi T$. We
 see that the interpolation regime at about $L\,T\approx 1$ is rather
 small and the asymptotic values are approached rapidly.

\subsection{Thermal \& volume dependence of the couplings}
The results obtained in the last Sections exemplified the thermal and
finite volume decays at the example of the thermodynamic observables,
the free energy density and the pressure. Both observables have power
counting dimension four, and indeed their flows scale with $k^4$ for
non-analytic regulators. This leads to non-trivial initial conditions
for these observables at the initial large cutoff scale
$k=\Lambda$. We expect only power-counting reduced problems for the
flow of the vertices, in particular for the mass $m_k^2$ and the
coupling $\lambda_k$. If this reduction leads to sufficiently strong
suppression in powers of $1/k$ even non-analytic regulators would
allow for a direct access to both the thermodynamics as well as the
flow of the couplings without resorting to non-trivial initial
conditions at least beyond leading order.

As for the thermodynamics it is illustrative to simply consider the
length derivative of the flows. For the effective potential this can
be directly read off from \eq{eq:dLG}, \eq{eq:derVfLPA},
\begin{align} 
  L\0{\partial \dot V_k}{\partial L} =-\012 \sumint_q
  \left( 2 \vec q^2 \partial_{\vec q^2} - L \0{\partial
      m_k^2}{\partial L} \partial_{m_k^2}\right) G_k\,\dot R_k\,. 
\label{eq:dLV}
\end{align} 

\begin{figure}[t]\begin{center}%
\includegraphics[width=\columnwidth]{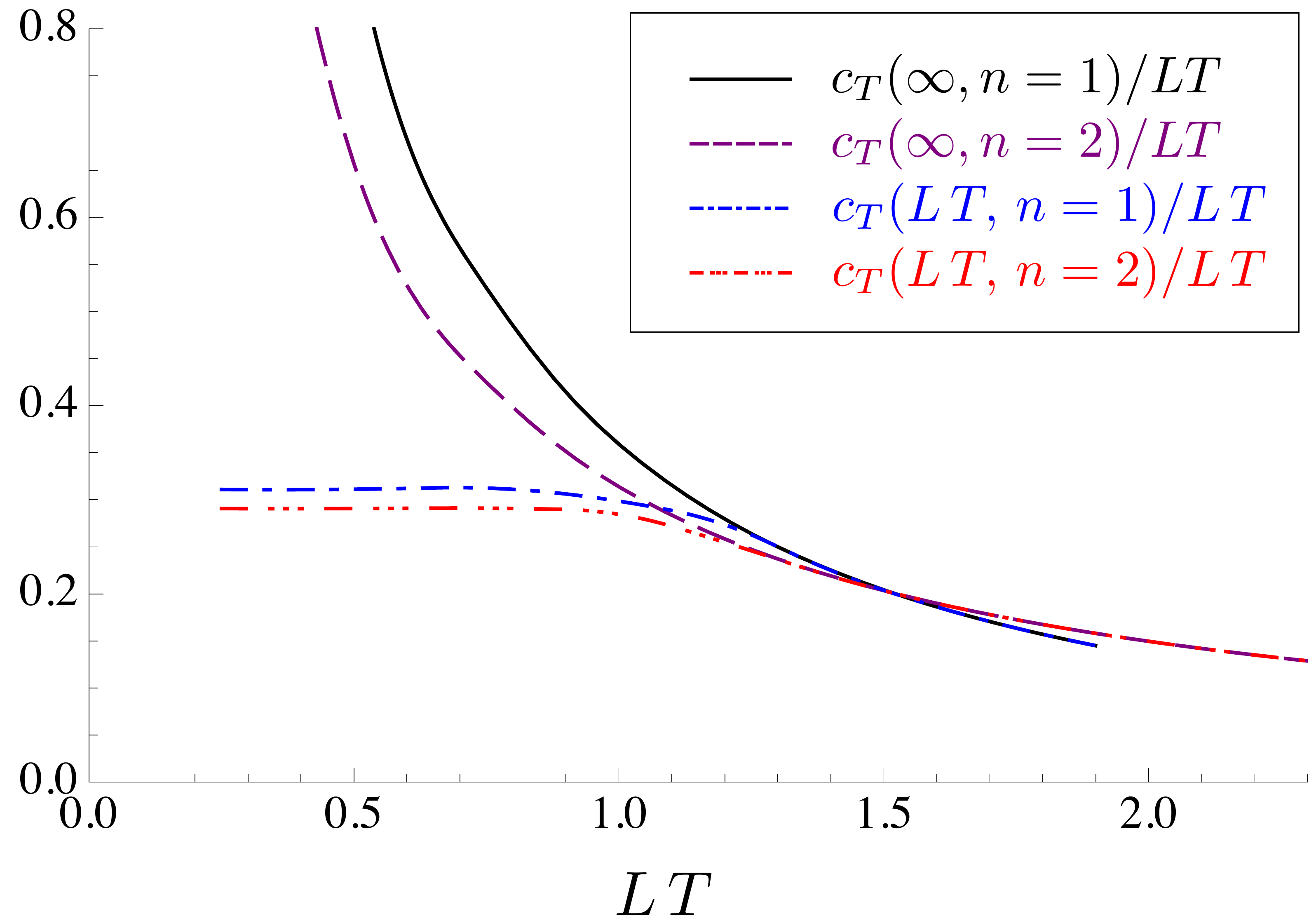}
\caption{Interpolation of $c_T$ for an exponential regulator,
  \eq{eq:Rexp}, with $m = 2$ from the asymptotic purely thermal decay
  to the asymptotic purely finite volume decay.}
\label{fig:cTVinterpolate}
\efc%

The second term in \eq{eq:dLV} can be viewed as an improvement
term. It does not dominate the flow. For the present purpose of a
power counting analysis we drop it and concentrate on the first
term. The flow of the effective potential reads for the
three-dimensional and four-dimensional flat regulators,
\begin{align} 
  L\0{\partial \dot V^{\text{\tiny{3dflat}}}_k(\rho)}{\partial L}
  \propto & \0{k^3}{L^3} \sum_{n_i\in \Z} \0{ \coth \0{k}{2 T}
    \sqrt{1+\0{m_k^2(\rho)}{k^2}} }{\sqrt{1+\0{m_k^2(\rho)}{k^2}}}
  \delta(k^2-\vec q^2)\,,
\end{align}
and 
\begin{align}
  L\0{\partial \dot V^{\text{\tiny{4dflat}}}_k(\rho)}{\partial L}
  \propto &2 \0{T}{L^3}\sum_{n_\mu\in \Z} \0{\vec
    q^2}{1+\0{m_k^2(\rho)}{k^2}} \delta(k^2-q^2)\,,
\label{eq:34dLV}\end{align}
respectively, where the superscripts indicate the chosen regulator.
The question of the sufficient decay that is at the heart of $T$ and
$L$-independent initial conditions can already be answered at one
loop. Thus, we integrate the flows in \eq{eq:34dLV} at one loop from
$k=0$ to $\Lambda$. This simply removed the $\delta$-function leading
to
\begin{align}\nonumber 
  V^{\text{\tiny{3dflat}}}_\Lambda(\rho)-V^{\text{\tiny{3dflat}}}_0(\rho)
  \propto & \016 \0{\Lambda}{ L^3} \sum_{\vec q^2\leq \Lambda^2} \0{ \coth
    \0{\Lambda}{2 T} \sqrt{1+\0{m_\Lambda^2(\rho)}{\Lambda^2}}
  }{\sqrt{1+\0{m_\Lambda^2(\rho)}{\Lambda^2}}}
  \,,\\[2ex]
  V^{\text{\tiny{4dflat}}}_\Lambda(\rho)-V^{\text{\tiny{4dflat}}}_0(\rho)\propto&
  \013 \0{T}{L^3}\sum_{q^2\leq \Lambda^2} \0{\vec
    q^2/\Lambda^2}{1+\0{m_\Lambda^2(\rho)}{\Lambda^2}}\,,
\label{eq:34dV}
\end{align}
where we have also integrated over the box size from infinite length
to $L$. The expressions for the one loop couplings $m_\Lambda^2$ and
$\lambda_\Lambda$ are derived from \eq{eq:34dV} by two and four
derivatives with respect to $\phi$, respectively. This involves terms
that are proportional to the first and second derivative of
\eq{eq:34dV} with respect to $m_\Lambda^2$, respectively. Hence, we
have additional suppressions with $1/\Lambda^2 $ and $1/\Lambda^4$,
respectively.  In turn, the sums in \eq{eq:34dV} grow proportional to
$\Lambda^3$ and $\Lambda^4$ on the average. Sweeping over a single
discrete frequency and spatial momentum value leads to a jump of the
effective potentials $V^{\text{\tiny{3dflat}}}_\Lambda$ and
$V^{\text{\tiny{4dflat}}}_\Lambda$ with a height proportional to
$\Lambda$ and $\Lambda^0$ respectively.  For the couplings this
entails a suppression of the single jumps with $1/\Lambda$ and
$1/\Lambda^2$ for $m_\Lambda^2$, and $1/\Lambda^3$ and $1/\Lambda^4$
for $\lambda_\Lambda$.

In the following, we restrict ourselves to the efficient regulator of
the form \eq{eq:Rexp} with $m=1$. For this choice the transition to
the vacuum theory happens at cutoff scales close to the first
frequency, and the cutoff scale can indeed be identified with a mass
scale $m_{\rm gap}$ of the order of the cutoff scale $k$. 

In principle, there are three different scales in the problem at hand:
vacuum mass, temperature and volume. In order to study the interplay
of temperature and finite volume only, we consider the massless theory
in the vacuum.  Note that this implies in general $m^2_k\neq 0$ for
cutoff scale $k\neq 0$. The vanishing mass at $k\rightarrow 0$
requires a fine tuning at $k=\Lambda$: for a given initial coupling
$\lambda_{\Lambda}$ we vary the initial value of the mass $m_\Lambda$
such that $m_{k\rightarrow 0}=0$. A generalisation to finite masses is
straightforward but not the main purpose of this work.

In summary, from the value $m_{k=0}$ we generate the initial condition
$m_\Lambda$ and $\lambda_\Lambda$ for mass and coupling flows,
\eq{eq:LPAobserve} with \eq{eq:loopfinV}, to be solved in a
self-consistent way. The functional behaviour of the coupling and,
consequently, of the mass is significantly different for large and
small values of $\lambda_\Lambda$.  For large initial values of the
coupling there are strong changes in mass and coupling, i.e.\ the running is
very strong. For smaller values of the coupling the running is slowed
down.  This results in the fact that the couplings at $k=0$ are much
closer to each other than at the initial scale. This behaviour is of
course well know and explained by the fact, that the coupling appears
quadratically and cubically, respectively, as a prefactor in the flows
of the mass and coupling.

\bfc%
\includegraphics[width=\columnwidth]{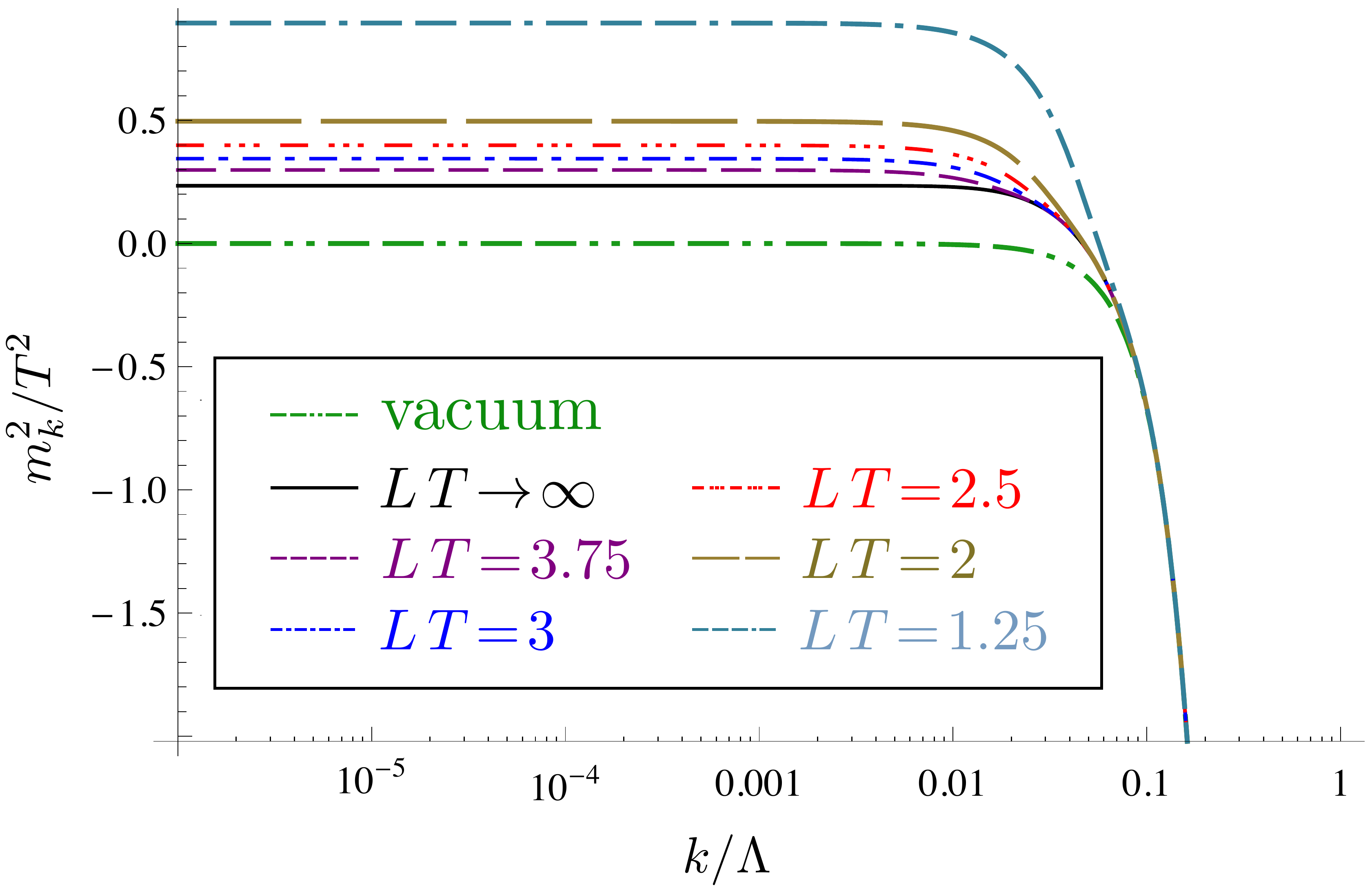}
\caption{Thermal and finite volume dependence of the mass for large UV
  coupling and $T/\Lambda=1/40$, and the exponential regulator, \eq{eq:Rexp} 
with $m=1$.}
\label{fig:mass_largecoupling}
\efc%

The effects of thermal fluctuations have been studied extensively in
e.g.\ \cite{Blaizot:2006rj, Blaizot:2010ut} for the $\phi^4$-theory,
for Yang-Mills theory this has been done in
\cite{Fister:2011uw,Fister:2011um,Fister:Diss}: As finite temperature
is an infrared modification of the theory, the theory is unchanged at
scales larger than the typical temperature scale, which is set by the
first Matsubara frequency $2\pi T$. Hence, above this scale the flow
follows the vacuum flow. Below this scale, temperature induces
additional fluctuations which accelerate the flow. For the mass this
results in the generation of a thermal mass in addition to the vacuum
mass.  However, once this generated scale is reached the mass
suppresses fluctuation in both flows of the mass and the coupling. As
a consequence, the flows are frozen below this scale and the values of
the coupling and mass saturate.

In this work, we consider the thermal field theory in a finite
volume. Similar to the compactification of the temporal direction due
to non-zero temperature, the finite edge length of the (cubic)
$3d$-box is a modification of the infrared. Therefore, its effect is
qualitatively similar to thermal effects.

\begin{figure}[t]\begin{center}%
\includegraphics[width=\columnwidth]{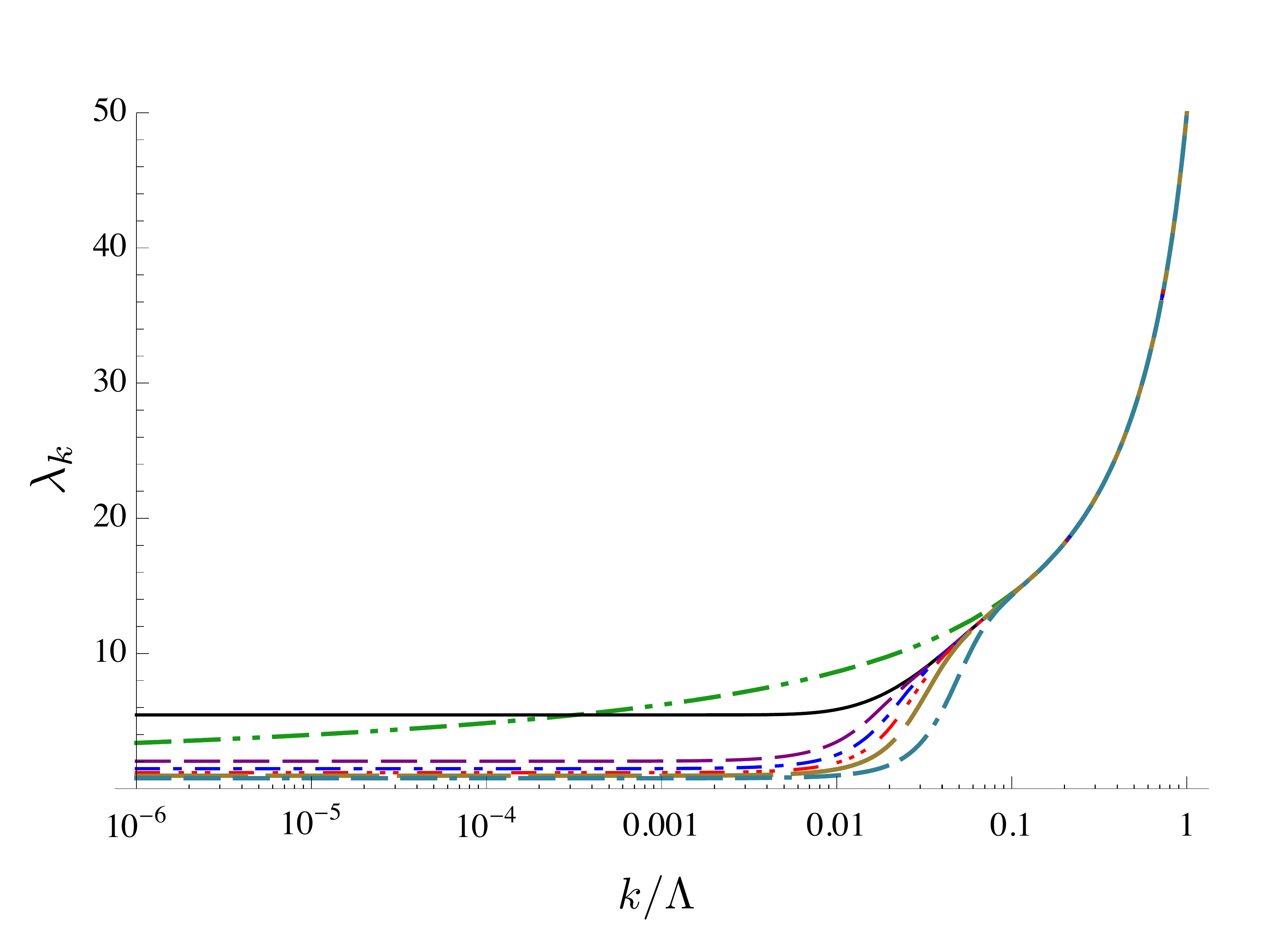}
\caption{Thermal and finite volume dependence for the coupling for
  large UV coupling. Regulator, parameters and colour-coding are the
  same as in \fig{fig:mass_largecoupling}.}
\label{fig:lambda_largecoupling}
\efc%
Once the theory is probed at
distances comparable to the volume fluctuations become stronger and
generate an additional mass.  At some point, however, the flows stop
as fluctuations are suppressed by the mass.  In general, it depends on
the magnitude of the three scales $L$, $T$ and $m_0$, as already the
largest scale effectively suppresses long range fluctuations and leads
to an independence of further macroscopic details. Having tuned to $m_0=0$,
the only relevant parameter that determines the behaviour of the
theory is the ratio of spatial length $L$ to temporal length $1/T$,
hence $L\, T$. Thus, as long as cutoff effects are correctly
taken care of, i.e.\ $\Lambda \gg 2\pi T$ and $\Lambda \gg 2\pi /L$,
see discussion above, different combinations of $L$ and $T$ give
similar results as long as their product is kept constant. In other
words, the effects of finite volume are always relative to the
temperature scale and it is thus sufficient to study only one
temperature for different lengths.

\begin{figure}[t]\begin{center}%
\includegraphics[width=\columnwidth]{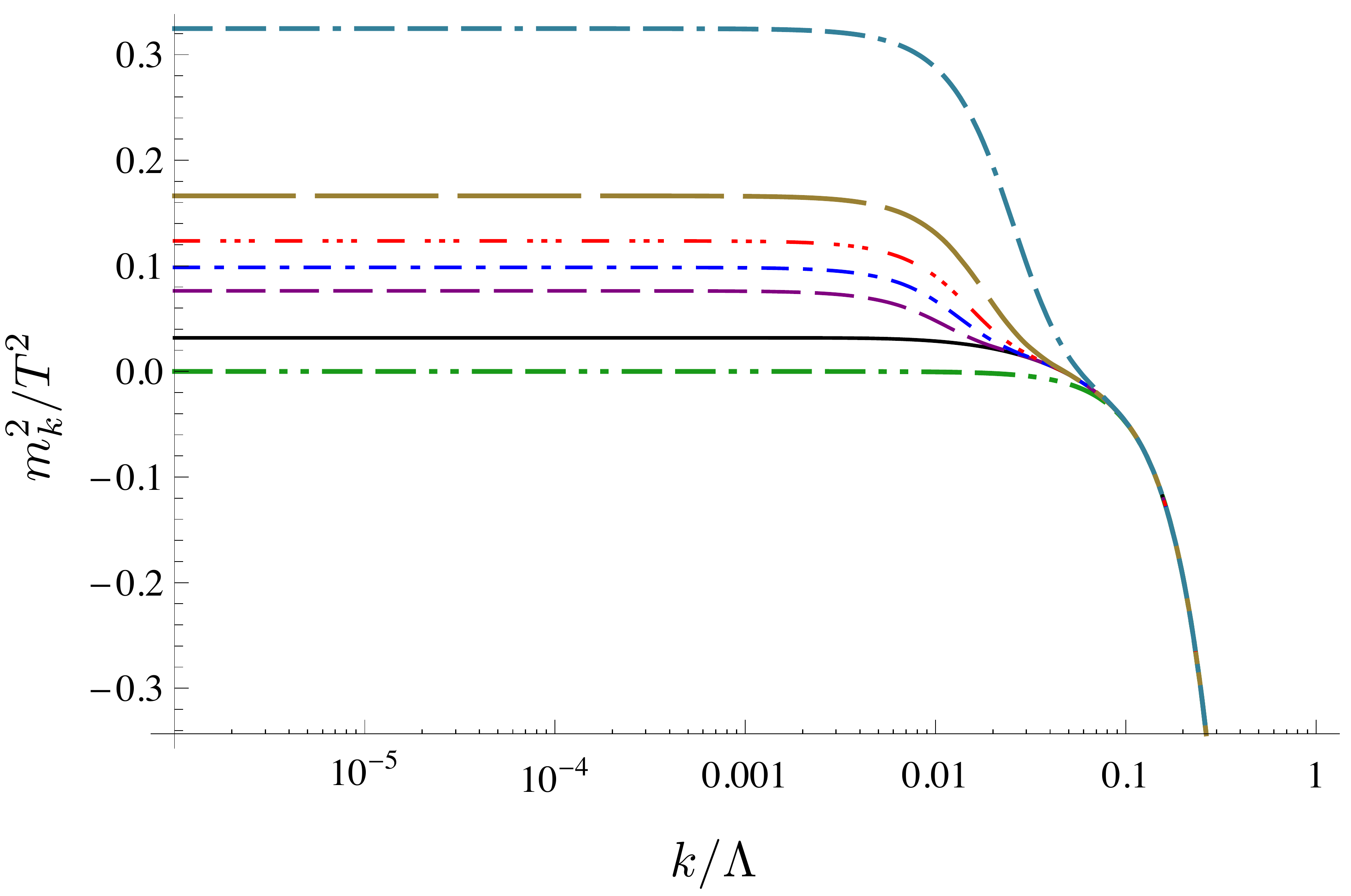}
\caption{Thermal and finite volume effects in the mass for small UV
  coupling. Regulator, parameters and colour-coding are the
  same as in \fig{fig:mass_largecoupling} but the coupling from \fig{fig:lambda_smallcoupling}.}
\label{fig:mass_smallcoupling}
\efc%
\begin{figure}[b]\begin{center}%
\includegraphics[width=\columnwidth]{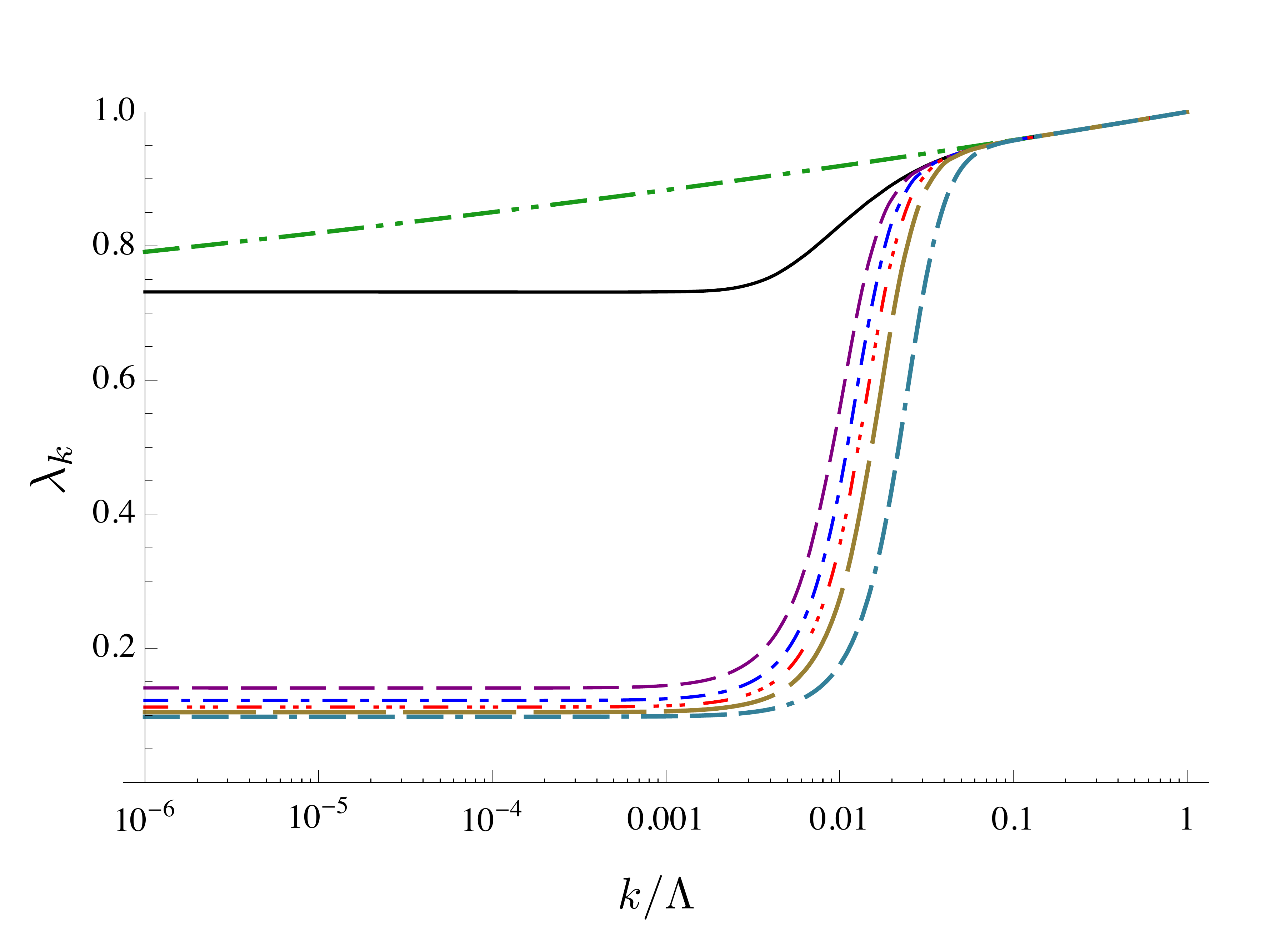}
\caption{Thermal and finite volume effects in the coupling for small
  UV coupling. Regulator, parameters and colour-coding are the
  same as in \fig{fig:mass_smallcoupling}.}
\label{fig:lambda_smallcoupling}
\efc%

In \fig{fig:mass_largecoupling}, \fig{fig:lambda_largecoupling} we
show cutoff scale dependence of the mass and the coupling for large
initial values of the coupling. In \fig{fig:mass_smallcoupling} and
\fig{fig:lambda_smallcoupling} we show the cutoff scale dependence for
small initial couplings. Note that the flows are damped by $k^2$,
unlike the flow of the potential.  Thus, the technical problem of 
"mode-counting" cutoff does affect the mass or coupling but becomes
only relevant at the level of the potential. It is clearly visible
that the effect of the finite volume is qualitatively equal to the
effect of finite temperature for both the mass,
cf. \fig{fig:mass_largecoupling} and \fig{fig:mass_smallcoupling}, and
the coupling, cf. \fig{fig:lambda_largecoupling} and
\fig{fig:lambda_smallcoupling}. Furthermore, the dependence on the
finite volume is monotonic in the sense that the larger the volume the
less additional flow is generated, i.e.\ the infinite volume limit is
approached. Interestingly, the limit of infinite volume is approached
very slowly, however, i.e.\ the value of $L\,T \gg 1$. This is due to
the fact that the finite volume affects three dimensions and is to be
compared with the compactification of the temporal direction only.
The mass flows for a massless and massive theory are
shown in \fig{fig:massive_flows}, which exemplifies that the vacuum
mass does not change the effect of finite temperature of volume
qualitatively. This justifies our choice $m_0=0$.

\begin{figure}[t]\begin{center}%
\includegraphics[width=\columnwidth]{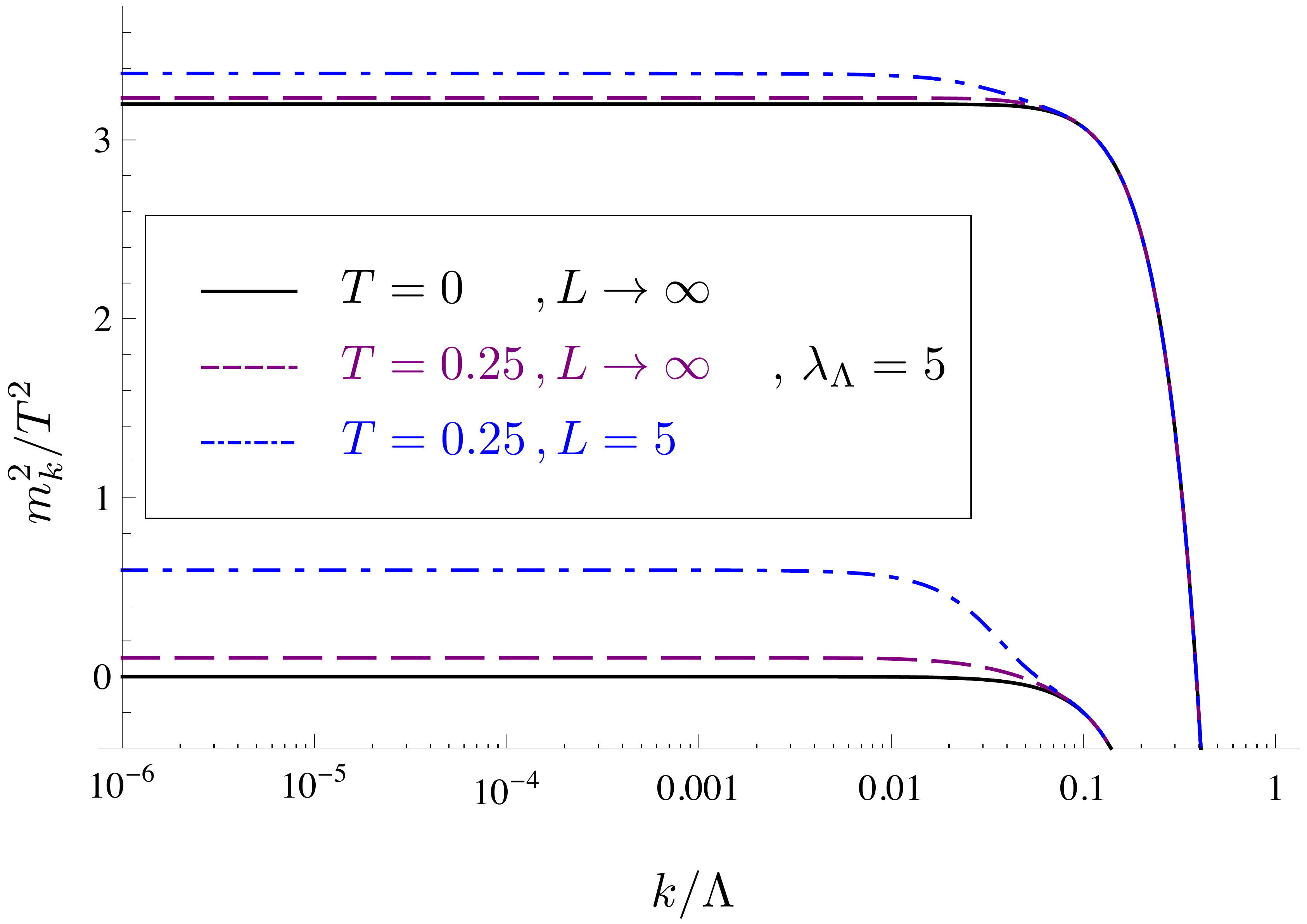}
\caption{Temperature and volume dependence of the mass for a
  massless (lower curves) and massive (upper curves) vacuum theory
  with $m^2_{k=0}=0.2$ for $\lambda_\Lambda=5$ at the initial scale. }
\label{fig:massive_flows}
\efc%
\begin{figure}[b]\begin{center}%
\includegraphics[width=\columnwidth]{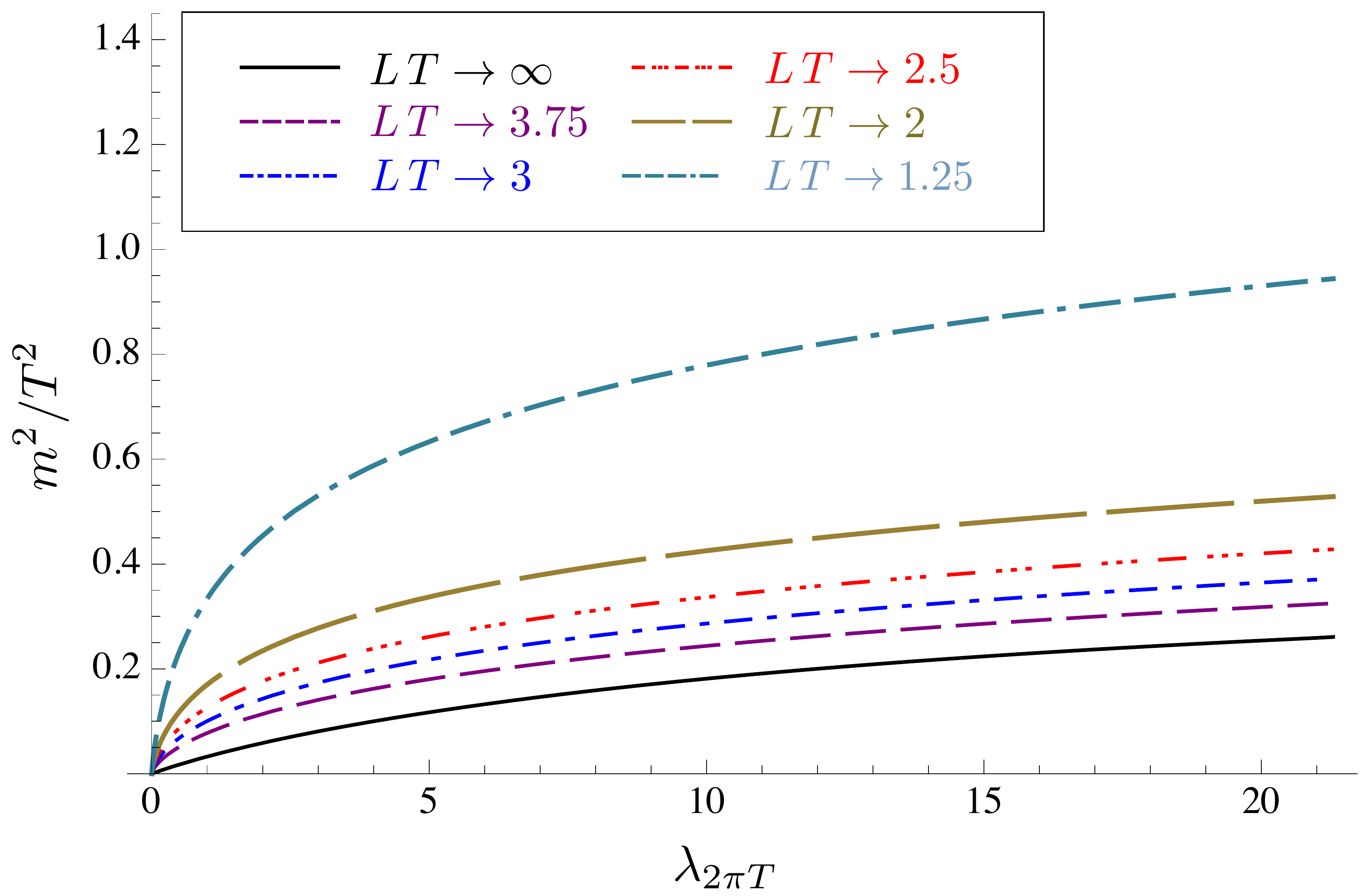}
\caption{Masses as a function of the coupling at $2\pi T$ for
  different volumes.}
\label{fig:masses_vs_coupling2piT}
\efc%
In \fig{fig:masses_vs_coupling2piT} we show the dimensionless mass as
a function of the value of the coupling at $2\pi T$. We choose this
value to facilitate comparison with other approaches where the
coupling is defined at that scale, e.g.\ perturbation theory. Again,
finite volume effects strengthen fluctuations and the effective mass
is therefore increased. Thus, the infinite volume limit for the mass
is approached from above. This is similar to the well-known thermal
behaviour, see \fig{fig:ratio_masses}.

Eventually, we consider the pressure. Note again that is can be evaluated
trivially by integrating the flow equation \eq{eq:floweq} with the
mass function computed above. In \fig{fig:dflengths} the flow of the
pressure is shown. Integrating the flow for a trivial initial
conditions gives the pressure. In \fig{fig:pressure} the pressure is
shown as a function of the value of the coupling at $2\pi T$ and
normalised to the Stefan--Boltzmann pressure. The infinite volume case
agrees with results in the literature. We see that both, increasing
the coupling and decreasing the volume leads to a smaller pressure.
Moreover, in comparison to the free energy density the thermal
pressure shows significantly less volume dependence. This reflects the
fact that it accounts for thermal fluctuations and their
volume-dependence.

\section{Conclusions}

We have investigated finite volume effects for the thermodynamics of a
$\phi^4$ theory within the functional renormalisation group
approach. We have focused on conceptual properties of the regulator
dependence of flows in finite volume compared to infinite volume. The
applicability of regulators can be decreased or even spoiled by the
non-analyticity of the regulator for momenta at the RG scale, $p^2=k^2$. We
find that the thermal flow of the effective action exhibits
oscillations along $k$. Their range becomes larger with the sharpness
of the regulator. In the limit of a sharp or flat regulator these
oscillations extend to infinity.
\begin{figure}[t]\begin{center}%
\includegraphics[width=\columnwidth]{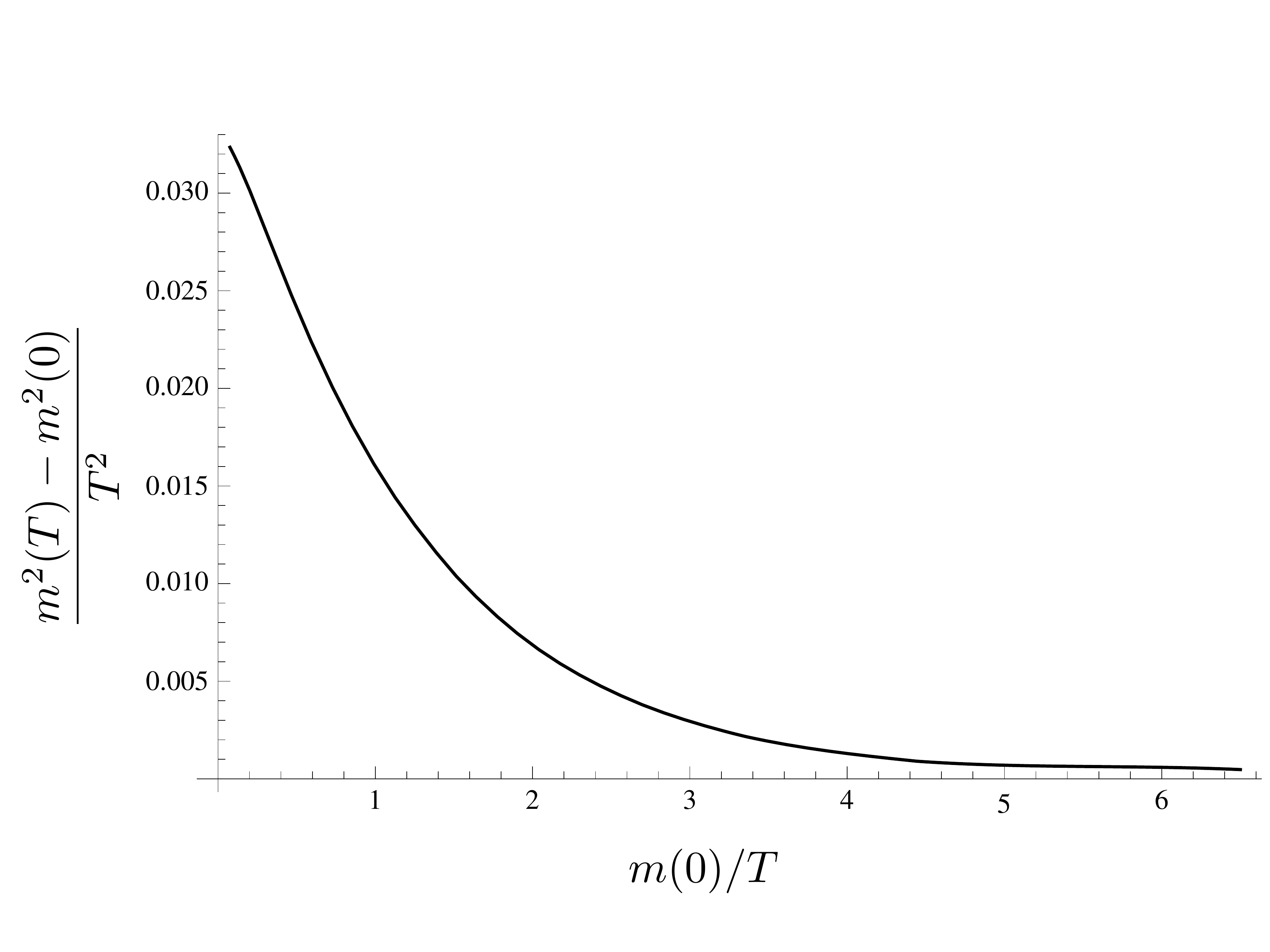}
\caption{Mass ratio of thermal and vacuum mass as a function of
  (fixed) vacuum mass over temperature. The ratio shows an exponential
  decay with $\exp -m(0)/T$.}
\label{fig:ratio_masses}
\end{center}
\end{figure}%
For a massless theory the only scales present are the temperature and
the spatial length. Hence, effectively the
theory only depends on one parameter, the ratio of lengths in
spatial to temporal directions, $L\, T$, where $L$ is the edge
length of a cubic box, and $1/T$ is the length in temporal direction.

The cutoff scale introduces an additional scale to the system and both
length scales are visible as onset scales for thermal and finite
volume effects respectively. For analytic regulators the onset scales
are proportional to $T$ and $L$, while the proportionality constant
depends on the chosen regulator.  Above the onset scales the flows
tend exponentially towards the corresponding vacuum flows. In turn,
for non-analytic regulators the exponential decay towards the vacuum
flows is missing, which makes quantitative studies of thermal and
finite volume effects cumbersome. 

\begin{figure}[t]\begin{center}%
\includegraphics[width=\columnwidth]{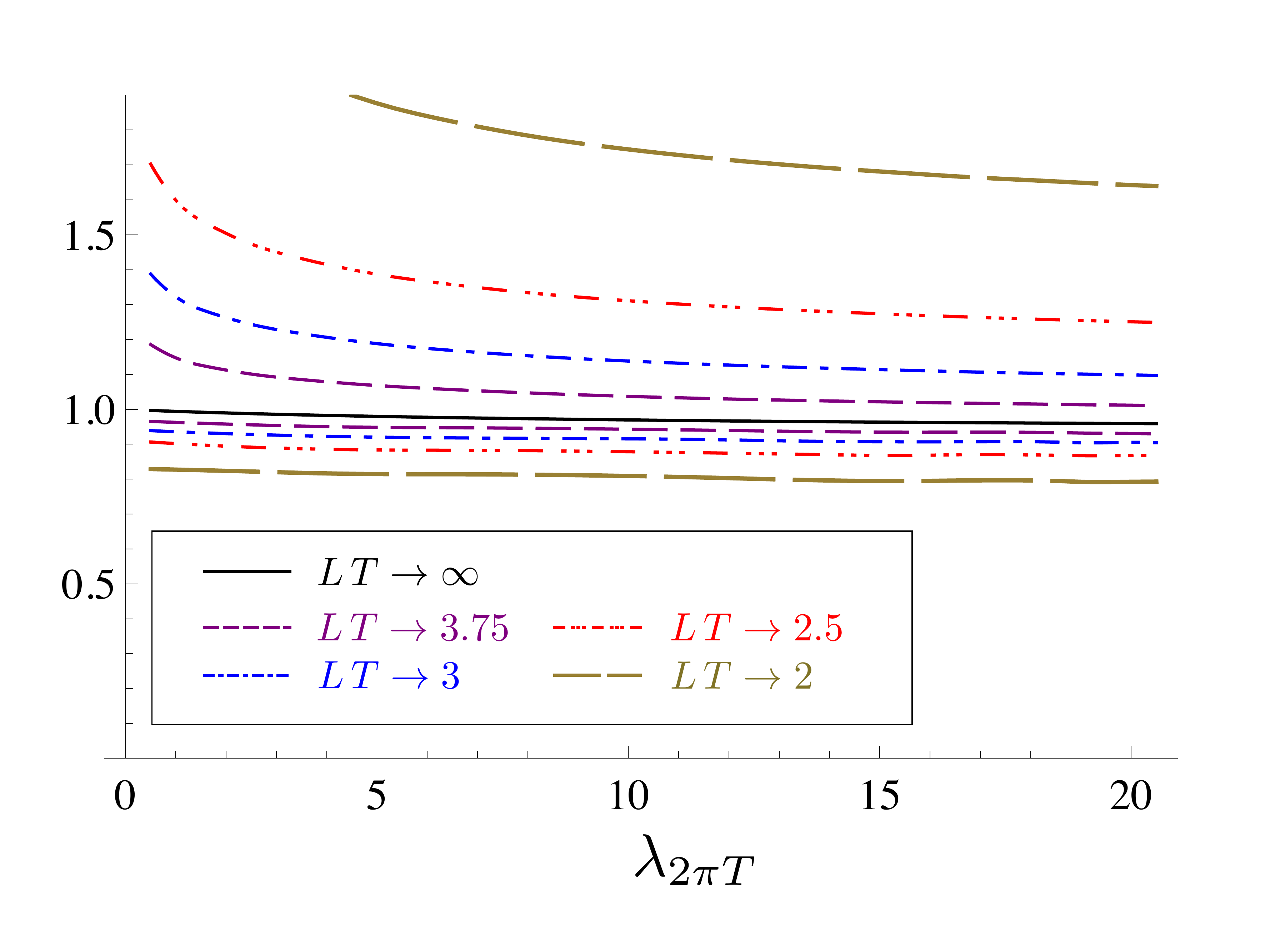}
\caption{Pressure, $p$, of $\phi^4$-theory in an infinite volume
  (black curve) and a finite volume (curves below infinite volume
  result) in comparison with the free energy, $-f_{k=0}$ (curves above
  infinite volume result). }
\label{fig:pressure}
\efc%

As expected, both finite volume and thermal effects show similar
dependences on the mass scales of the theory including the cutoff
scale. This is directly seen in volume dependence of the thermal mass
already, where the mass is larger for smaller volumes. As a
consequence, the infinite volume limit of the mass is reached from
above.

Furthermore, we have studied the pressure.  The infinite volume limit
is reached quicker for larger couplings, because correlations are
transferred in a more efficient way and fluctuations are washed out
before the length scale is reached.  Therefore, for smaller couplings
the effects of the finite length is stronger. This is again a
manifestation that these are pure infrared effects.

The structural aspects of the finite volume dependence studied in the
present work are quite general.  The present findings carry over
straightforwardly to $O(N)$ models with more than one field component,
to phenomenological low energy systems such as the quark--meson model,
and, to some extend to full QCD. In the latter case the gauge sector
requires some special attention. This will be studied elsewhere.
\\

{\it Acknowledgments} --- We thank J.~Braun and B.-J.~Schaefer for
discussions. LF is supported by ERC-AD-267258. Moreover, this work is
supported by the Helmholtz Alliance HA216/EMMI and by ERC-AdG-290623.\\[1ex]

\appendix
\section{Pressure flows for a flat box cutoff}\label{app:flatbox}
\begin{figure}[t]\begin{center}%
\includegraphics[width=\columnwidth]{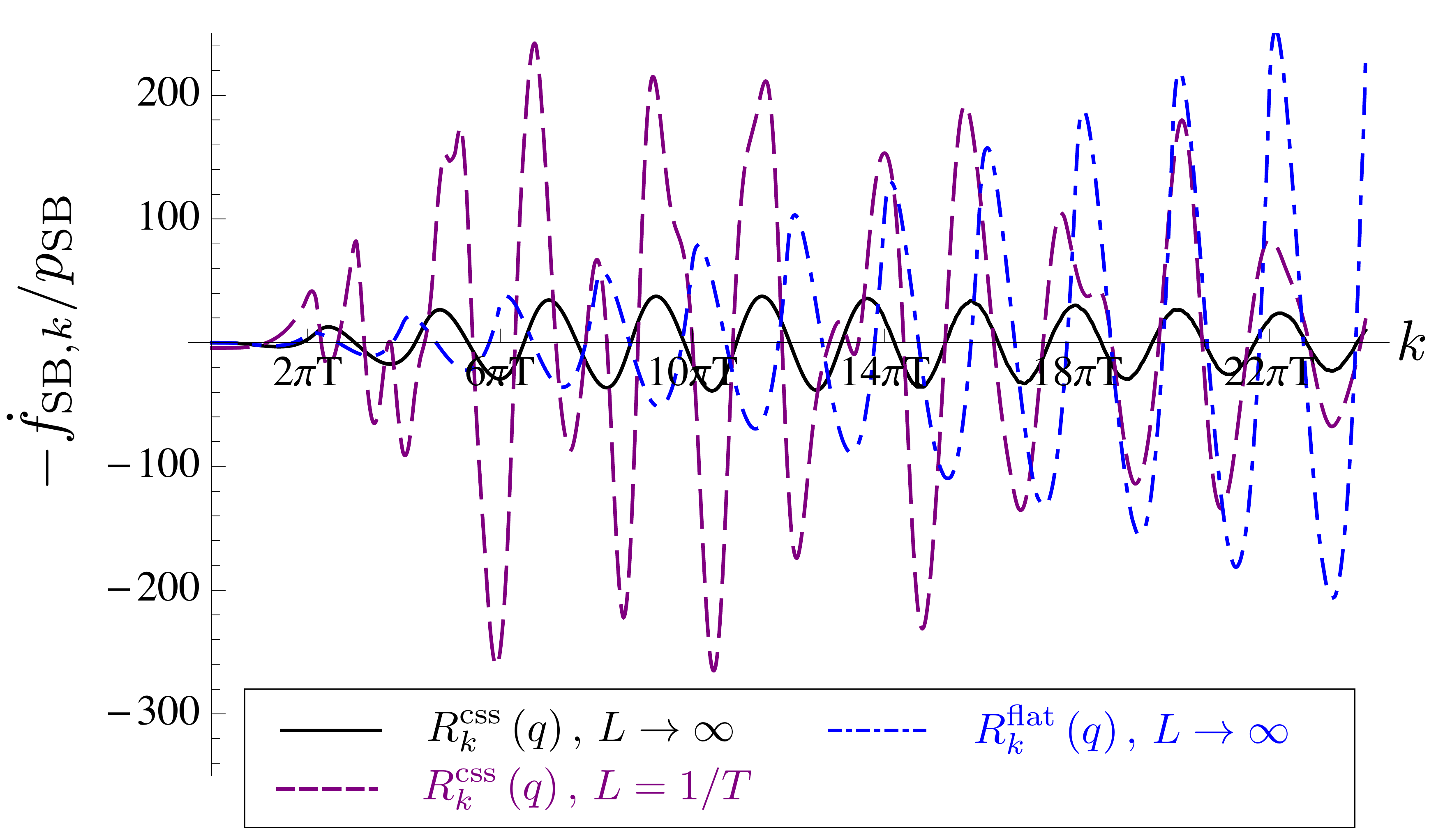}
\caption{Flow of the Stefan--Boltzmann free energy
  $\dot{f}_{\text{\tiny{SB}},k}$ for the css regulators \eq{eq:css} in
  finite and infinite volume compared to the flat regulator
  \eq{eq:Rflat}, in the infinite volume.}
\label{fig:css}
\efc
For a fully analytical approach we choose the product cutoff 
\begin{align}\label{eq:prodRk}
R_k^{\rm prod}(q)= \prod_{\mu=0}^4 R_k^{\text{\tiny{flat}}}(q_\mu)\,, 
\end{align}
with 
\begin{align}\label{eq:prodGk}
G_k^{\rm prod}(q)= \0{1}{k^2}\prod_{\mu=0}^4 \theta(k^2-q_\mu^2) +
\0{1}{q^2}\left[1- \prod_{\mu=0}^4 \theta(k^2-q_\mu^2)\right]\,, 
\end{align}
The flow of $f$ is then given by 
\begin{widetext} 
\begin{align}
  \dot{f}_{\text{\tiny{SB}},k} = \0{T}{\CV}\Biggl\{ (2 N_T +1)\left(
    \0{N_L}{3}+N_L^2 +\0{2 N_L^3}{3}\right)^3 + 3 (2 N_L+1) \left(
    \0{N_T}{3}+N_T^2 +\0{2 N_T^3}{3}\right)\left( \0{N_L}{3}+N_L^2
    +\0{2 N_L^3}{3}\right)^2\Biggr\} -\0{ 8 k^4}{ 27 \pi^4}\,,
 \label{eq:SB_flowprodfV}\end{align}%
\end{widetext}
with 
\begin{align}\label{eq:NL} 
  N_L=\left[ \0{k\,L}{2 \pi} \right]\,,\qquad N_T=\left[ \0{k}{2 \pi
      T} \right]\,,
\end{align}
where the brackets indicate the largest integer number smaller than
the argument.  The enveloping function of \eq{eq:SB_flowprodfV} rising
as $k^3$. This exemplifies the problem related to non-analytic
regulators.

\section{Compactly supported smooth regulator}\label{app:css}
Here we test the decay for compactly supported smooth (css) regulators
suggested in \cite{Nandori:2012tc}, \beqa%
R_k^{\rm css}(q)&=&q^2 r_k^{\rm css}\left(q^2/k^2\right)\,,\nn\\[2ex]
r_k^{\rm css}(y)&=&\0{\exp\left\{ c y_0^b/\left(1-y_0^b\right)-1
  \right\}} {\exp\left\{ c y^b/\left(1-y^b\right)-1 \right\}}
\Theta\left(1-y\right)\,,
\label{eq:css}
\eeqa%
with the normalisation $y_0$ such that $r_k^{\rm css}(y_0)\equiv
1$. We choose the parameters $b=1,\, c=1$ and $y_0 =1/2$ leading to a
standard exponential decay. In \fig{fig:css} we show the flow of the
Stefan--Boltzmann free energy $\dot{f}_{\text{\tiny{SB}},k}$ for the
css regulators \eq{eq:css} in comparison to the flat regulator
\eq{eq:Rflat}. In both cases of infinite and finite volume
$\dot{f}_{\text{\tiny{SB}},k}$ does not show the exponential decay for
large cutoff scales $k$.

\section{Flows for the $L$-derivatives of the
  couplings}\label{app:derV}
In the present $\phi^4$-approximation to the LPA one has to compute
$m_k^2(L)$ and $\lambda_k(L)$. More generally, in a polynomial
expansion of the LPA, where
\begin{align}\label{eq:vecl}
  \vec \lambda=(\lambda_1,\lambda_2,\lambda_3,...)\,,\quad {\rm
    with}\quad \lambda_1=\kappa\,,\quad \lambda_2 =\lambda\,,
\end{align}
there are additional couplings. Their flows are given schematically by
\begin{align} 
\dot \kappa =& - \012 \int_q G_k(q,\rho_0)\,\dot R_k(q)\,
\end{align}
and
\begin{align}
  \dot \lambda_n -\dot \kappa \lambda_{n+1} =& \012 \int_q
  G_k^{(n)}(q,\rho_0)\,\dot R_k(q)\,,
\label{eq:flown}\end{align} 
where $G^{(n)}(q,\rho)=\partial_\rho^nG(q,\rho)$, using
\eq{eq:dtV}. Taking the $L$-derivative (or $\CV$-derivative) of
\eq{eq:flown}, 
with 
\begin{align}\label{eq:derVl}
  \lambda_n'=L \partial_L \lambda_n = 3 \CV \0{\partial
    \lambda_n}{\partial \CV}\,,
\end{align} 
leads to flow equations for the $L$-derivatives of
$\lambda_n$, therefore,
\begin{align}
\partial_t  \kappa'  = -\012 \int_q \left( 2 \vec
    q^2 \partial_{\vec q^2} -
    \lambda_i' \partial_{\lambda_i} \right)
  G^2_k(q,\rho_0)\,\dot R_k(q)\,,
\label{eq:flowkappadV}\end{align} 
and 
\begin{align}\nonumber 
  & \hspace{-.5cm}\partial_t \lambda_n' - \partial_t \lambda_1'
  \lambda_{n+1}+ \partial_t \lambda_1 \lambda_{n+1}'\\[2ex]
  =&\hspace{.5cm} -\012 \int_q \left( 2 \vec q^2 \partial_{\vec q^2} -
    \lambda_i' \partial_{\lambda_i} \right) G^{(n)}(q,\rho_0)\,\dot
  R_k(q)\,.
\label{eq:flowndV}\end{align} 
Note that $\rho_0=\max(0,\kappa)$ depends on $\lambda_1 =\kappa$.
Eqs.~\eq{eq:flowkappadV},\eq{eq:flowndV} can be solved on the basis of
a solution for $\lambda_k$ in a given volume $\CV$ (a cube with box
length $L$), and a regulator that guarantees the decay of volume
fluctuations for $k\to\infty$, that is
\begin{align}\label{eq:flucV0}
\lim_{k\to\infty} \lambda_n' =0\,.
\end{align}
\Eq{eq:flucV0} requires sufficiently smooth regulators in the spatial
momentum directions. Moreover, iterations of this procedure leads to
flow equations of $N$th $L$-derivatives of the couplings $\vec
\lambda$ in the basis of $j=0,...,N-1$ $L$-derivatives of $\vec
\lambda$ and their flows.

The present derivations trivially extend to the situation of different
spatial lengths $L_i$ for $i=1,2,3$. We add that the present procedure
applies more generally to flows of derivatives with respect to
external parameters such as volume, temperature, chemical potential,
and couplings.

\bibliography{finV}
\end{document}